\documentclass[10pt,
twocolumn,
 amsmath,amssymb,
 aps,
prx,longbibliography
]{revtex4-1}

\usepackage[dvipdfmx]{graphicx}% Include figure files
\usepackage{dcolumn}% Align table columns on decimal point
\usepackage{bm}% bold math
\usepackage{braket}
\usepackage{color}
\usepackage[normalem]{ulem}
\newcommand{\ve}[1]{\mathbf{#1}}
\newcommand{\txr}[1]{\textcolor{black}{#1}}

\newcommand{\txb}[1]{\textcolor{black}{#1}}
\definecolor{darkgreen}{rgb}{0,0.6,0}
\newcommand{\txg}[1]{\textcolor{black}{#1}}
\definecolor{purple}{rgb}{0.5,0,0.5}

\newcommand{\mi}[1]{\textcolor{black}{#1}}
\newcommand{\minw}[1]{\textcolor{black}{#1}}

\begin{document}

%\preprint{APS/123-QED}

\title{%\txr{Mott and antiferromagnetic}  transitions of anisotropic two-dimensional Hubbard model \txr{with emergence of 
Quantum criticality of bandwidth-controlled Mott transition}
%}% Force line breaks with \\
\author{Kensaku Takai}
\affiliation{Department of Applied Physics, University of Tokyo, 7-3-1 Hongo, Bunkyo-ku, Tokyo 113-8656 }
%\author{Marcin Raczcowski}
\author{Youhei Yamaji}
\affiliation{Center for Green Research on Energy and Environmental Materials, National Institute for Materials Science, Namiki, Tsukuba-shi, Ibaraki, 305-0044, Japan}
\author{Fakher F.  Assaad}
\affiliation{Institut f\"{u}r Theoretische Physik und Astrophysik and W\"{u}rzburg-Dresden Cluster of Excellence ct.qmat,
Universit\"{a}t W\"{u}rzburg, Am Hubland, D-97074 W\"{u}rzburg, Germany}
\author{Masatoshi Imada}
\affiliation{Toyota Physical and Chemical Research Institute, 41-1 Yokomichi, Nagakute, Aichi, 480-1192, Japan,}
\affiliation{Research Institute for Science and Engineering, Waseda University, 3-4-1 Okubo, Shinjuku-ku, Tokyo, 169-8555, Japan,} 
\affiliation{Department of Engineering and Applied Sciences, Sophia University, 7-1 Kioi-cho, Chiyoda, Tokyo 102-8554, Japan}
%\affiliation{%
% Department of Applied Physics, University of Tokyo, 7-3-1 Hongo, Bunkyo-ku, Tokyo 113-8656
%}%



\begin{abstract}
Metallic states near the Mott insulator show a variety of quantum phases including various magnetic, charge ordered states and high-temperature superconductivity in
various transition metal oxides and organic solids.
The emergence of a variety of phases
and their competitions are likely intimately associated with quantum transitions between the electron-correlation driven Mott insulator and metals characterized by its criticality, and is related to many central questions of condensed matter.  The quantum criticality is, however, not well understood when the transition is controlled 
by the bandwidth through physical parameters such as pressure.  Here, we quantitatively estimate the universality class of the transition characterized by a comprehensive set of critical exponents  
by using a variational Monte Carlo method implemented as an open-source innovated quantum many-body solver,  
with the help of established scaling laws
at a typical bandwidth-controlled Mott transition.  
The criticality indicates a weaker charge and density instability in contrast to the filling-controlled transition realized by carrier doping, implying a weaker instability to superconductivity as well.    The present comprehensive clarification opens up a number of routes for quantitative experimental studies for complete  understanding of  elusive quantum Mott transition and nearby strange metal that cultivate future design of functionality.  
\end{abstract}

\maketitle

%\tableofcontents
\section{Introduction}
The Mott transition is a metal-insulator transition driven by the Coulomb repulsion of electrons in crystalline solids. %without necessarily accompanying spontaneous symmetry breaking. 
It is driven either by controlling the ratio of the interaction strength to the bandwidth (bandwidth-controlled transition) or by carrier doping to the Mott insulator (filling-controlled transition)~\cite{RevModPhys.70.1039}. 
The two types of control are widely realized in
%often accompanied by the first-order jump of physical quantities or the emergence of the charge inhomogeneity in strongly correlated electron systems such as 
organic solids~\cite{RevModPhys.89.025003,KatoReview2014} and transition metal compounds~\cite{RevModPhys.70.1039}.

The filling-controlled transition has been relatively well studied motivated by the high temperature superconductivity in the cuprates. Theoretically estimated criticality of the Mott transition was suggested to cause the charge instability that gives birth to severe competitions of the high temperature superconductivity, strange metal, antiferromagnetism, nematicity and charge inhomogeneity including charge order in the cuprates~\cite{misawa2014superconductivity,ImadaSuzuki2019,ImadaReview2021}.
%Experimentally observed charge inhomogeneity 
It is also understood from the tendency towards the first-order transition that generates a miscibility gap in the carrier density near the Mott insulator.  When the first-order transition %and miscibility gap 
can be suppressed, criticality emerges around the marginal quantum critical point (MQCP)~\cite{PhysRevB.72.075113}.
%as the fundamental origin of the charge instability. 
\mi{The MQCP critical exponents have not been well explored in experiments, partly because various competing phases including superconductivity and effect of disorder preempt or mask criticality. However, the emergence of exotic phases including the superconductivity in the cuprates may be governed by the underlying MQCP and therefore the understanding of the MQCP has crucial importance to reveal the mechanism of the competing phases.}  
%\cite{Kn,Knphys,PhysRevB.75.115121,PhysRevB.72.075113, uniImada}. 

On the other hand, 
%strongly correlated electron systems such as transition metal oxides and organic solids often show 
the bandwidth-controlled transitions have also been widely observed. 
%in organic solids~\cite{RevModPhys.89.025003,KatoReview2014} and transition metal compounds~\cite{RevModPhys.70.1039}.  
They normally appear as first-order transitions, which terminate at a critical endpoint at nonzero temperatures. 
The universality class of this endpoint %at nonzero temperatures 
was proposed to belong to that of the classical Ising-model%as gas-liquid transitions
~\cite{PhysRevLett.43.1957,Limelette89}. 
When the critical temperature %of the Mott critical endpoints 
is reduced to zero as the MQCP,  the universality class should be distinct~\cite{PhysRevB.75.115121,PhysRevB.72.075113}. %, uniImada}.  
One of the central questions is whether the universality class can lead to strong quantum fluctuations and quantum entanglement, which triggers emergence of novel functionality including high-temperature superconductivity \mi{similarly to the incentive to gain insights for the filling-controlled case~\cite{ImadaReview2021}}.  However, %the quantum criticality of 
the bandwidth-controlled Mott transition at the MQCP and the related charge instability are not well explored even theoretically.  
%The most striking difference of the quantum Mott transition from that at finite temperatures was emphasized \txrs{to arise from}\txr{by} the role  played by the Fermi surface. 
%While the Fermi surface is ill defined at finite temperatures and expected to be irrelevant to the criticality, the \txr{clearly defined} Fermi surface necessarily vanishes at the quantum Mott transitions and likely governs the universality class at zero temperature. 
%However, it still remains unclear how the Fermi surface vanishes and affects the criticality.

\begin{figure}[h!]
		\begin{center}
			\includegraphics[width=6cm,clip]{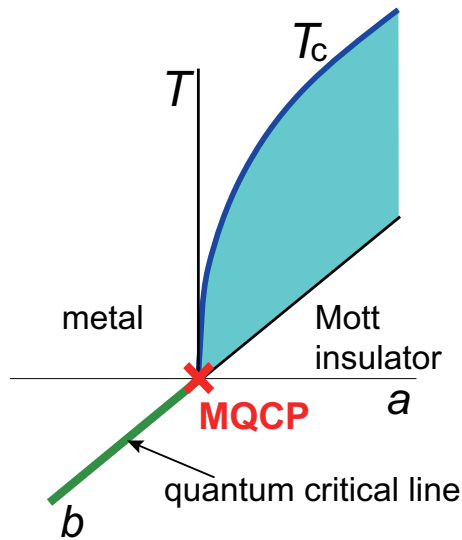}   
		\end{center}
	\caption{\mi{Schematic phase diagram of Mott metal-insulator transition. The MQCP (red cross) is the quantum critical point between the metal and the Mott insulator and simultaneously the end point of the first order transition and the quantum critical line (green line). Finite temperature critical point ($T=T_{\rm c}$) (dark blue curve) appears as the endpoint of the first order boundary (light blue shaded surface.  $a$ and $b$ represent the control parameters and are given by a combination of $t_{\perp}$ and $U$ in the present case. In the bandwidth-control case in general, the electron filling is fixed at an odd integer in this whole $T$-$a$-$b$ parameter space. For details see Ref.\onlinecite{PhysRevB.75.115121}.}}
	\label{MQCPSchematic}
	\end{figure}

\mi{We summarize the basic structure around the MQCP of the metal insulator transition found in the earlier work, which is illustrated in Fig.~\ref{MQCPSchematic}~\cite{PhysRevB.75.115121}. The MQCP appears as the endpoint of the finite temperature critical line, \minw{namely, the endpoint of the first-order transition}, while it also appears as the endpoint of the quantum critical line (QCL) running at temperature $T=0$. The reason why the critical line continues beyond the MQCP is that the metal and insulator must always have a clear phase transition boundary at $T=0$ unlike the case of the quantum Ising model such as that with the transverse magnetic field where the transition disappears beyond the conventional quantum critical point. Our focus in this paper is the universalty class of the bandwidth-controlled MQCP and not the criticality of the QCL, because the MQCP is excpected to show stronger quantum fluctuations and entanglement with enhanced charge fluctuations that may trigger exotic phases~\cite{PhysRevB.75.115121}.} 

\minw{In the literature, the motivation of the study on the quantum critical point (QCP) in general has come from the expectations for novel physics, where finite critical temperature is lowered to zero and associated diverging quantum fluctuations emerge, which may induce exotic phases. In the present case, this corresponds to the MQCP appearing as a single point at $T=0$, although the distinction between the MQCP and QCL is not well appreciated in the literature. The reason may be due to the fact that the QCL does not exist in the conventional critical point (QCP) arising from symmetry-breaking transitions.
%However, MQCP is different from the conventional QCP because of the extension of the QCL as is seen in Fig.~\ref{MQCPSchematic}. 
Along the quantum critical line (QCL), the criticality should be different from the MQCP in general.}

\minw{%A big research trend to understand the conventional QCP occuring at a single point has partly been motivated by the expectation for the instability induced by quantum critical fluctuations toward exotic phases. 
%MQCP appears in the same way as QCP as a single point at $T=0$.  
Significance of the QCP including the MQCP is that the first-order transition starts from the QCP, which opens the possibility of coupling to divergent zero-wavenumber modes. In the case of the metal-insulator transition, this appears as the divergent charge fluctuations.
On the other hand, the QCL exists even in the noninteracting case as in the simple band-insulator metal transition. 
%In fact, in the case of the filling control transition, the nonsingular features of QCL were already clarified in Ref.~\onlinecite{PhysRevB.75.115121}. 
For instance, in Ref.~\onlinecite{PhysRevB.75.115121}, the criticality of the MQCP was clarified for the filling-controlled transition in detail and the critical exponents are identified as $\alpha= -1, \beta = 1, \gamma= 1, \delta= 2, \nu=1/2$ and $\eta=0$, where $\gamma= 1$, and $\delta= 2$ lead to the divergence of the charge compressibility $\kappa \propto 1/x$, where $x$ is the doping concentration.  The divergent compressibility at the MQCP was supported in a 2D Hubbard model study~\cite{misawa2014superconductivity}. In contrast, $\alpha= 0, \beta = 1, \gamma= 0, \delta= 1, \nu=1/2$ and $\eta=0$ were reported for the QCL. Here, the exponents $\alpha= 0, \gamma= 0$, and $\delta= 1$ imply that the fluctuations are not diverging. This is because of the absence of the opening of the first-order transition and indeed it is equivalent to the band-insulator-to-metal transition in usual noninteracting systems. The divergent charge fluctuations for the filling-controlled MQCP on the verge of the phase separation or the charge inhomogeneity opens the possibility of emergent exotic phases such as unconventional superconductivity associated with this divergence and fluctuations. In the dynamical mean field theory (DMFT) calculation, the metal-insulator critical point appears at a finite temperature, at which it was shown that the charge compressibility diverges~\cite{PhysRevLett.89.046401}.  However, in the DMFT, one cannot lower the critical temperature to zero to reach the MQCP, while in 2D one can see such an evolution to the MQCP. Therefore, it is natural to pose a question how the interplay between the diverging charge fluctuation and quantum fluctuations takes place at the MQCP for the bandwidth-controlled case in 2D. In other words, the nontriviality of the MQCP lies in the fact that the first-order metal-insulator transition and the resultant MQCP does not exist in the non-interacting case and it is purely the interaction effect. 
By considering this background and the significance with a direct connection to the quantum critical phenomena in general, we study the MQCP rather than the QCL. 
}

  In this article, we study the mechanism and criticality of the bandwidth-controlled quantum Mott transitions. 
For this purpose, we employ  anisotropic two-dimensional Hubbard models %with geometrical frustrations 
at half filling as a typical example. 
We study the model by using
a state-of-the-art variational Monte Carlo method (VMC)~\cite{JPSJ.77.114701,PhysRevB.90.115137}, where the open source code is available~\cite{Misawa_mVMC}. 
See Sec.~\ref{Method} A for details of the numerical method. 
The solution of the model shows the existence of the MQCP.
We estimate a comprehensive set of critical exponents of the MQCP, which shows a perfect consistency with the scaling theory, which indicates a weaker charge and density instability in contrast to the filling-controlled transition by carrier doping, implying a weaker instability to superconductivity as well.   Since the earlier experimental as well as theoretical studies by the dynamical mean-field study suggest the exponents different from the present results, we discuss the origin of the discrepancy. 

This paper is organized as follows: In Sec.\ref{Model}, we introduce the model. In Sec.\ref{Phase Diagram}, the phase diagram is shown in the plane of the Hamiltonian parameters, which reveals the MQCP. In Sec.~\ref{Estimate of MQCP}, the critical exponents of the MQCP are thoroughly estimated. In Sec.~\ref{Scaling Analysis}, the estimated exponents are analyzed in terms of the scaling theory. Section~\ref{Discussion and Summary} is devoted to Discussions and Summary.

\section{Model} \label{Model}

For the purpose of clarifying the generic feature of the bandwidth-controlled Mott transition, as an example, we study the $t$-$t_{\perp}$-$t'$ Hubbard model at half filling defined by the following Hamiltonian :
\begin{eqnarray}
\hat{H}=&-&t\sum_{\langle i,j\rangle_x,\sigma}\hat{c}^{\dag}_{i\sigma}\hat{c}_{j\sigma}
-t_{\perp}\sum_{\langle i,j\rangle_y,\sigma}\hat{c}^{\dag}_{i\sigma}\hat{c}_{j\sigma}\nonumber\\
&+&t'\sum_{\langle\langle i,j\rangle\rangle,\sigma}\hat{c}^{\dag}_{i\sigma}\hat{c}_{j\sigma}
+U\sum_{i}\hat{n}_{i\uparrow}\hat{n}_{i\downarrow},
\label{aniso2DHubbard}
\end{eqnarray}  
 where $\hat{c}_{i\sigma}$ ($\hat{c}_{i\sigma}^{\dagger}$) annihilates (creates) a spin-$\sigma$ electron at site $i$ and $\hat{n}_{i\sigma}$ is its number operator. Here, $t$ ($t_{\perp}$) is the hopping between the nearest-neighbor sites in the $x$-($y$-) direction, $t'$ is that between the next-nearest-neighbor sites and $U$ represents the on-site Coulomb repulsion. %(See inset of Fig.~\ref{fig:phase} for the lattice geometry.)
 The lattice structure of the present model is depicted in the inset of Fig.~\ref{fig:phase}, where the intra-chain transfer $t$ and inter-chain transfer $t_{\perp}$ constituting the square lattice are geometrically frustrated with the next-nearest-neighbor transfer $t'$. % fixed at $t/2$.
The onsite Coulomb interaction $U$ monitors the correlation effects and the control of $U/t$ triggers the bandwidth-controlled Mott transition. 
%See Supplementary Information (SI) A for the shape of the Fermi surface at $U=0$ for the present model.  
%We further discuss a universal feature of the bandwidth-controlled Mott criticality beyond specific models. 
 In this model, by taking the nearest neighbor transfer $t$ along the chain direction as the energy unit, namely $t=1$, the interchain hopping $t_{\perp}$ acts as the parameter to control the dimensionality between 1D ($t_{\perp}=0$) and 2D ($t_{\perp}=t$), which enables the control of the Mott transition temperature to zero, namely allows us to study the MQCP. Here we fix the ratio of the next nearest neighbor hopping $t'$ to $t_{\perp}$ as $t'=t_{\perp}/2$.

\mi{Although we  employ a specific model, the notion of  universality  that  characterizes  the  2D MQCP,  renders  the  details of the model irrelevant.   
%It merely has  to share the  symmetries of   the MQCP  and  of  the universality class of the bandwidth control MQCP in 2D may belong to the same universality irrespective of details of the model and parameters, 
The MQCP essentially emerges between the metal and Mott insulator and it appears as the endpoint of both of the first-order transition and the continuous quantum critical line as sketched schematically in Fig.~\ref{MQCPSchematic}. In addition it does not retain the C$_4$ rotational symmetry, which is common to the experimental structure in the organic solids ~\cite{RevModPhys.89.025003,KatoReview2014} and offer the possibility to capture the generic feature of the 2D MQCP. }
\minw{Although the transfer terms introduce slightly 1D-like anisotropy, we confirm that spin and charge fluctuations show isotropic singular behavior below and represents a typical 2D criticality.}
%by analyzing the critical exponents defined in Methods C in detail. 
We obtain a comprehensive set of critical exponents that are consistent with each other in light of the scaling theory. In contrast to previous theoretical and experimental studies at finite temperatures $T>0$ above the classical critical endpoint \txg{to infer a zero-temperature exponent}~\cite{PhysRevLett.107.026401,FurukawaKanoda2015}, we focus on the quantum case directly at $T=0$.
% The mVMC method has been developed to accurately obtain ground state wave functions of strongly correlated systems without the negative sign problem. 
% We apply \txr{the} mVMC to the anisotropic Hubbard model on a square lattice and determine the ground-state phase diagram. 
%Here the \txr{isotropic} two-dimensional square lattice with the geometrical frustration term $t'$ is formed for $t=t_{\perp}$, while the lattice becomes quasi-one-dimensional for small $t_{\perp}$. 
%Here, we set $t'=t_{\perp}/2$ to study more frustrated case because we can approach by mVMC without the negative sign problem (see Fig. \ref{fig:fs} (a)). 
%By introducing the stronger frustration $t'$, we expect the realization of metal-insulator transitions without any symmetry breaking.
We show, in Supplementary Materials (SM) A, the Fermi surface for the noninteracting case.
It changes from 1D-like open Fermi surface for small $t_{\perp}$ to 2D-like closed one by increasing $t_{\perp}$ separated by the Lifshitz transition at $t_{\perp}\approx 0.62$.
Similar models have been studied before~\cite{PhysRevB.65.115117,PhysRevB.84.045112,PhysRevLett.116.086403}. 
Here we focus on the criticality of the Mott transition, for which we assume that the universality class does not depend on the details of the model.

\section{Phase Diagram} \label{Phase Diagram}
\vspace{5mm}
%%%%%%%%%%%%%%%%%%%%%%%%%%%%%%%%%%%%%%%%%%%%%%%%%%%%
%    Figure 1
%%%%%%%%%%%%%%%%%%%%%%%%%%%%%%%%%%%%%%%%%%%%%%%%%%%%

%\section{\Large Phase Diagram}

\begin{figure*}
\begin{center}
\includegraphics[width=0.8\textwidth]{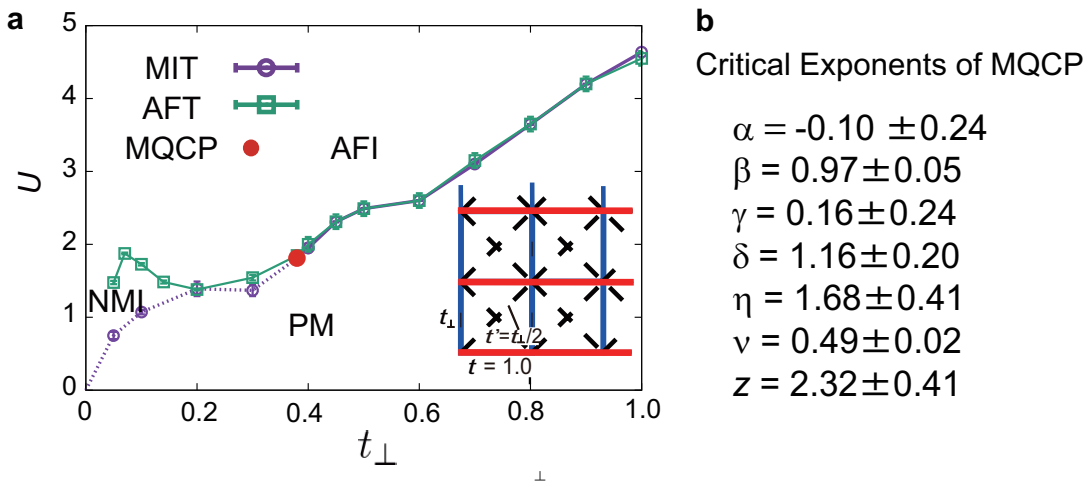}
\end{center}
\caption{ 
{\bf a} Ground-state phase diagram obtained by the present VMC calculation. 
The purple solid and broken lines with open circles indicate the first-order and continuous metal-insulator transition (MIT) boundaries, respectively.
The green solid curve with open squares is for the antiferromagnetic transitions (AFT) (see Sec.~\ref{Method} D for the method to determine the MIT and see section F of SM for the AFT). Red large circle depicts the MQCP. 
Error bars are determined by considering the errors of size extrapolations and statistical errors of Monte Carlo calculations for finite-size systems.
Inset: Lattice structure used for the present study; $t$-$t_{\perp}$-$t'$ Hubbard model with the nearest neighbor intrachain (red bonds), interchain (blue bonds) and next-nearest-neighbor (broken black bonds) hoppings $t, t_{\perp}$ and $t'=t_{\perp}/2$, respectively. We take $t$ as the energy unit.  \\%\txb{$\rightarrow$This figure is updated.}}
{\bf b} Critical exponents of MQCP estmated in this article.} 
\label{fig:phase}
\end{figure*}

We first summarize the obtained ground-state phase diagram of the metal, insulator and magnetic phases separated by metal-insulator and antiferromagnetic transitions %for the anisotropic Hubbard model (\ref{aniso2DHubbard}) 
in the parameter space of $U$ and $t_{\perp}$ in Fig. \ref{fig:phase}.
%, we show the phase diagram containing the metal-insulator and antiferromagnetic phase boundaries. 
\txr{Hereafter, we mainly focus on the metal-insulator transition. (Although we do not discuss details, the antiferromagnetic transition is discussed in Secs.F and I of SM).
For details of the method to determine the phase boundary, see Sec.~\ref{Method} D.}
 The transition is of first-order for large $t_{\perp}$ with
a jump in physical quantities %including opening/closing of the charge gap 
while it changes to a continuous one for smaller $t_{\perp}$ detected only by the continuous opening/closing of the charge gap (see SM, Sec. B). The first-order and continuous transitions meet at the MQCP.
For the first-order part, the transition temperature as well as the 2D Ising nature of the transition vanishes   %and the critical temperature 
%which reduces to zero 
at the MQCP.  %By monitoring $t_{\perp}$, 
We find the MQCP roughly around $t_{\perp}=t^{\rm MQCP}_{\perp}\sim 0.4$ and $U=U^{\rm MQCP}\sim 1.8$, %, which allows us to study its universality class in the later part of this paper. 
%More precise determination of the MQCP 
which will be more precisely estimated in the later part of this article. 
For $t_{\perp}>t^{\rm MQCP}_{\perp}$, magnetic and metal-insulator transitions occur essentially simultaneously as a first-order transition. On the other hand, for $t_{\perp}<t^{\rm MQCP}_{\perp}$, %we find that 
the two transitions become separated (see Sec. F of SM for the magnetic transition) and a nonmagnetic insulator (NMI) phase emerges, %in the insulator side
but we do not go into details of the NMI and leave it for studies elsewhere. 
\mi{We also do not study the universality of the quantum critical line depicted as the purple dotted line in Fig.~\ref{fig:phase}.}
\mi{Although the metal-insulator and antiferromagnetic transitions look slightly separated even for $0.2<t_{\perp}<t_{\perp}^{\rm MQCP}$, we do not exclude the possibility of a simultaneous transition within the numerical error bar.}
The overall phase structure obtained %from the VMC method 
here is essentially similar to that obtained by the cluster dynamical mean field theory (CDMFT) at low temperature~\cite{PhysRevB.103.125137}.
A small kink-like structure of the phase boundary around $t_{\perp}\sim 0.6$ is related to the Lisfshitz transition in the corresponding noninteracting model (see SM, Sec. A for the Fermi surface of the case $U=0$).

%%%%%%%%%%%%%%%%%%%%%%%%%%%%%%%%%%%%%%%%%%%%%%%%%%%%
%    Figure 2
%%%%%%%%%%%%%%%%%%%%%%%%%%%%%%%%%%%%%%%%%%%%%%%%%%%%
%%%%%%%%%%%%%%%%%%%%%%%%%%%%%%%%%%%%%%%%%%%%%%%%%%%%

%%%%%%%%%%%%%%%%%%%%%%%%%%%%%%%%%%%%%%%%%%
\section{Estimate of MQCP and its critical exponents} \label{Estimate of MQCP}
\vspace{5mm}
%%%%%%%%%%%%%%%%%%%%%%%%%%%%%%%%%%%%%%%%%%%

We now present our numerical results on the universality class 
%of the bandwidth controlled Mott transition 
at the MQCP.
% in the present example of the anisotropic 2D Hubbard model.
% obtained from the VMC calculation.
See  Sec.~\ref{Method} C for definitions of the critical exponents, $\alpha, \beta, \gamma, \delta, \nu, z$ and $\eta$ analyzed below. 
\mi{Since we need to estimate the position of the MQCP first and the MQCP is defined by the point where the first-order transition disappears, we first estimate when the jump of physical quantities characteristic of the first-order transition vanishes. The conventional scaling analysis does not work accurately unless the MQCP point is precisely estimated.
}%Comprehensive clarification of the exponents is crucially important to understand full nature of the Mott transition. %~\ref{CriticalExponent}. 
%As is determined in the last section, our MQCP was estimated roughly at $t_{\perp}=t_{\perp c}=0.38$ and $U=U_c=1.8$.

%%%%%%%%%%%%%%%%%%%%%%%%%%%%
%\subsection{Estimate of MQCP and its critical exponents}
%%%%%%%%%%%%%%%%%%%%%%%%%%%%%

\begin{figure}[h!]
\begin{center}
\includegraphics[width=0.45\textwidth]{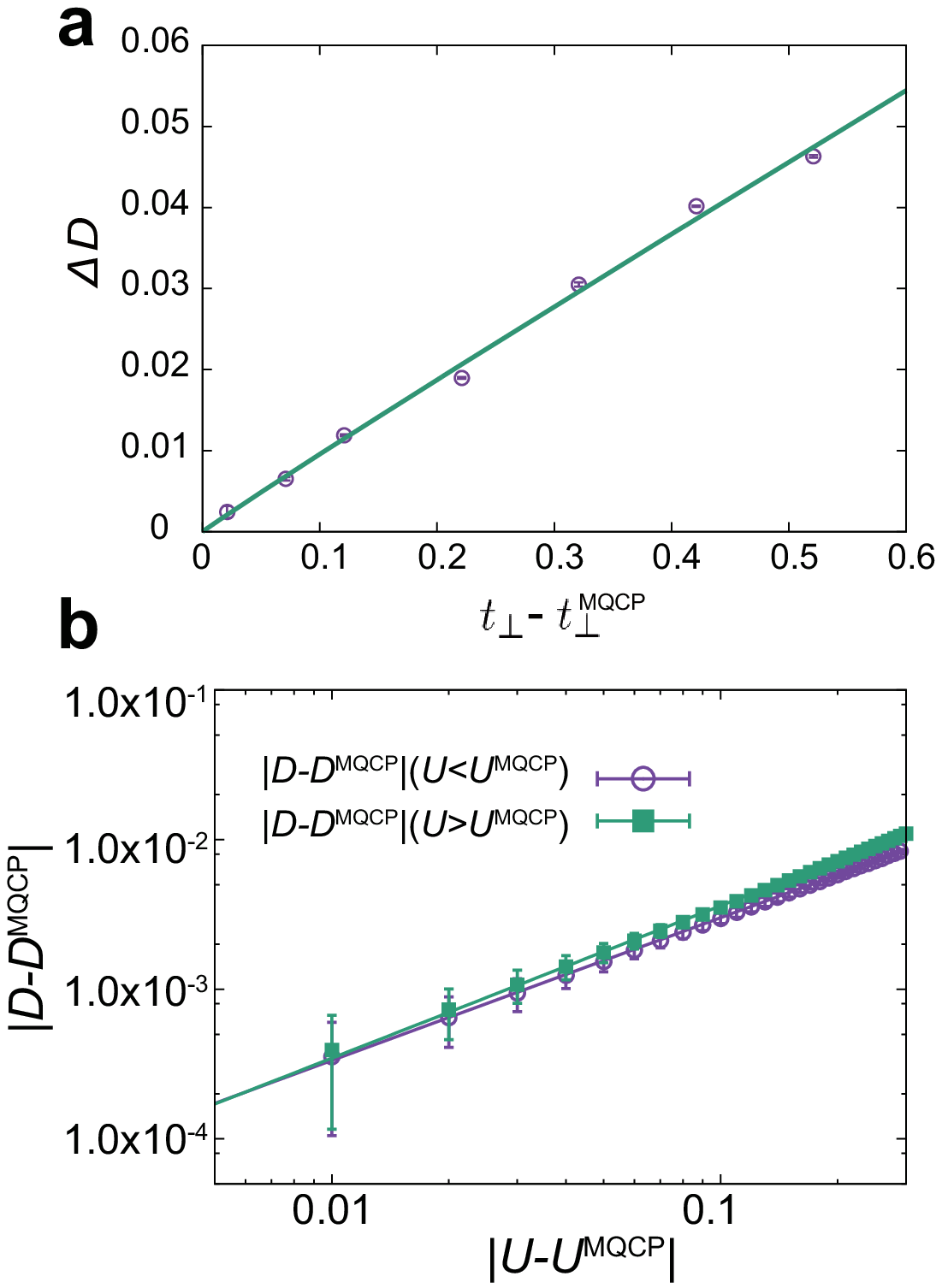}
\end{center}
\caption{
%\txb{[If possible, these figure will be shown with CDMFT's results.]}
{\bf a} $t_{\perp}$-dependence of
 %$\chi^2$-value for the fittings for $f(t_{\perp})$ (inset) and
 jumps of extrapolated double occupancy $\Delta D$ with the choice of $t_{\perp}^{\rm MQCP}=0.38$ determined in Methods D. %***\ref{MQCPchi2}. 
Green curve represents the optimized fitting leading to $\beta=0.97\pm 0.05$.
{\bf b} $U$ dependence of
 extrapolated double occupancy $D$ at $t^{\rm MQCP}_{\perp}=0.38$. %, and $|U-U^{\rm MQCP}|$ dependence of $D$ \txb{at $t^{\rm MQCP}_{\perp}=0.38$}.
 Green (purple) line represents the fittngs to estimate $\delta_{\rm I}$ for $U>U^{\rm MQCP}$ ($\delta_{\rm M}$ for $U<U^{\rm MQCP}$), which indicate $\delta_{\rm I}= 0.98 \pm 0.03$ and $\delta_{\rm M}= 1.05 \pm 0.04$.
%\txb{[***Two figures are merged.]
}
%\txm{The fitting curve must reach the origin. Is it really 0.97 power?}
%
\label{fig:dD_exp}
\end{figure}

The critical exponent $\beta$ of the MQCP (Eq. (\ref{beta})) is estimated from the jump of the double occupancy of electrons on the same site, $\Delta D=D_{\rm metal}-D_{\rm ins}$, where the double occupancy in the metallic (insulator) side is $D_{\rm metal}$ ($D_{\rm ins}$) along the first-order transition line in the region $t_{\perp}>t_{\perp}^{\rm MQCP}$ (see Eq.(\ref{Double_Occupancy}) for the definition of the double occupancy). 
\if0
to the functional form
\begin{align}
\Delta D(t_{\perp})=a |t_{\perp}-t_{\perp}^{\rm MQCP}|^{\beta}
\label{beta}
\end{align}
near the MQCP, where $a$, $\beta$ and $t_{\perp}^{\rm MQCP}$ are fitting parameters.
\fi
The fitting of the VMC numerical data in the range  $0.4 \le t_{\perp} \le 0.9$ plotted in Fig.~\ref{fig:dD_exp}{\bf a} shows that
the mean squared error 
by defining
the
%\txcs{$\chi^2$
%\txc{normal} 
%distribution around} 
mean given by
%\txcs{form} 
Eq.~(\ref{beta}) becomes the minimum when we employ the MQCP point at $t_{\perp}^{\rm MQCP}\sim0.38 \pm 0.05$ and $\beta =0.97\pm 0.05$ as is shown in Methods C and D (Fig.~\ref{chi_tp-tpc}{\bf a}).
%for the $t_{\perp}$ dependence %after optimizing $a$ and $\beta$ for each $t_{\perp}$ 
%as is shown in Fig.~\ref{fig:chi2}a in Methods and 
The green curve in Fig.~\ref{fig:dD_exp}{\bf a} is the resultant optimized fitting.
%the inset of Fig.~\ref{DeltaD_tperp} supporting that the MQCP is located most likely at $t_{\perp}^{\rm MQCP}\sim0.38 \pm 0.05$. 
%By considering the error bar, we employ $t_{\perp c}=0.38$.
%The \txr{optimized} fitting curve around the MQCP is shown as the green solid line in Fig. \ref{DeltaD_tperp}
%indicating $\beta =0.97\pm 0.05$. 
The error bar for $t_{\perp}^{\rm MQCP}=0.38$ estimated by the bootstrap method (see  Sec.~\ref{Method} E for details of the bootstrap) is included in the error bar of $\beta$.
The estimated $\beta$ is similar to $\beta=1$ in the filling-controlled transition predicted in the literature~\cite{PhysRevB.72.075113}.

\begin{figure}[h!]
\begin{center}
\includegraphics[width=0.45\textwidth]{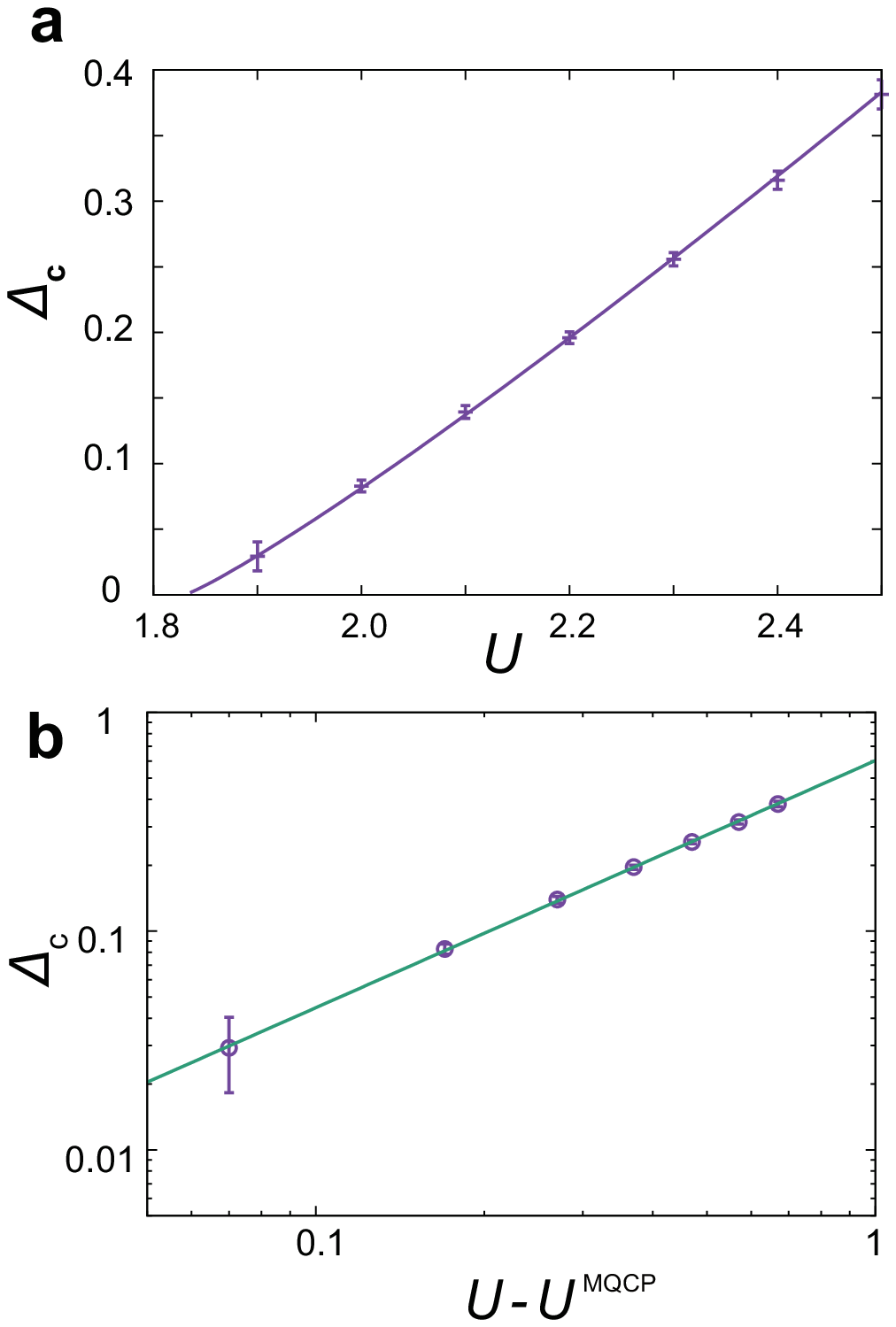}
\end{center}
\caption{\label{fig:tp0.38}The exponent $\nu z$ estimated from the charge gap $\Delta_{\rm c}$ at MQCP. $U$ dependence of $\Delta_{\rm c}$ in the linear plot ({\bf a}) and the logarithmic scaling plot ({\bf b}) for $28\times 28$ site around the  MQCP ($t_{\perp}^{\rm MQCP}=0.38$ and $U^{\rm MQCP}=1.83$).
%(b) $U-U^{\rm MQCP}$-dependence of $\chi^2$-value for the fittings for $f(U-U^{\rm MQCP})$ (incerted figure) and charge gaps $\Delta$.
Purple curve in {\bf a} and green line in {\bf b} show the same fitting by the estimate of  the MQCP point and the critical exponents, indicating $\nu z=1.13 \pm 0.19$.}
\label{Deltac_U-Uc}
\end{figure}

We also simultaneously determine the critical value of $U$ at the MQCP and critical exponents $\delta$ and $\nu z$ by the combined analysis with Eq.(\ref{chi2_U}) at $t_{\perp}^{\rm MQCP}=0.38$, and obtain $U^{\rm MQCP}=1.83 \pm 0.03$, $\nu z=1.13\pm 0.19$, 
%. \txb{The critical exponent $\delta$ defined in Methods C (Eq.~(\ref{delta})) from the $U$ dependence of $D$ is estimated separately in the insulating and in the metallic phases as  
$\delta_{\rm I}= 0.98 \pm 0.03$ and $\delta_{\rm M}= 1.05 \pm 0.04$ (see Figs.~\ref{fig:dD_exp}{\bf b} and \ref{Deltac_U-Uc} as well as Methods C and D),
where $\delta$ is estimated separately in the insulating ($\delta_{\rm I}$) and in the metallic ($\delta_{\rm M}$) phases.

These results imply that the nonsingular linear term proportional to $|U-U^{\rm MQCP}|$ makes the precise estimate of $\delta$ difficult, if $\delta \le 1$.  However, we will clarify that $\delta\sim 1.0$ is consistent with other scaling analyses.
%\txb{[Prof. Imada's comments: This paragraph must be refined later. Detailed analyses and comparisons with other (earlier) theories/experiments should be extensively added here or in Discussion section.$\rightarrow$ After the data for doublons in the AFI side will be polished with reduced error bars until next or next next Skype, discussions will be refined.]}
The exponent is the same again with the filling-controlled MQCP estimated  as $\delta=1$ in Refs.~\onlinecite{PhysRevB.75.115121} and \onlinecite{PhysRevB.72.075113} within the statistical error.
%%%%%%%%%%%%%%%%%%%%%%%%%%%%%%%%%%%%%%%%%%%%%%%%%%%%
%    Figure 3
%%%%%%%%%%%%%%%%%%%%%%%%%%%%%%%%%%%%%%%%%%%%%%%%%%%%
%%%%%%%%%%%%%%%%%%%%%%%%%%%%%%%%%%%%%%%%%%%%%%%%%%%%

%%%%%%%%%%%%%%%%%%%%%%%%%%%%%%%%%%%%%%%%%%%%%%%%%%%%
%    Figure 4
%%%%%%%%%%%%%%%%%%%%%%%%%%%%%%%%%%%%%%%%%%%%%%%%%%%%
%%%%%%%%%%%%%%%%%%%%%%%%%%%%%%%%%%%%%%%%%%%%%%%%%%%%

%%%%%%%%%%%%%%%%%%%%%%%%%%%%%%%%%%%%%%%%%%%%%%%%%
\section{Scaling Analysis} \label{Scaling Analysis}
\vspace{5mm}
%%%%%%%%%%%%%%%%%%%%%%%%%%%%%%%%%%%%%%%%%%

In our calculation, we obtained $\beta\sim 1.0$, $\nu z\sim 1.1$ and $\delta\sim 1.0$.
We now analyze this result in the framework of scaling theory.
Here, the singular part of the ground-state energy $E$ around the MQCP satisfies the form 
\begin{eqnarray}
E\propto \xi^{-(d+z)},
\label{Ehyperscaling}
\end{eqnarray}
where $\xi$ is the unique length scale that diverges at the MQCP, and 
$d$ and $z$ are the spatial dimension and the dynamical exponent, respectively. 
This scaling theory was examined in Ref.~\onlinecite{PhysRevB.75.115121}, where critical exponents 
satisfy the following scaling relations: 
%( see Sec.~\ref{Method} C for the definition of the critical exponents):%\ref{CriticalExponent} *** 
%for the summary of the scaling hypothesis and derivation):
\begin{eqnarray}
\label{scaling_relation}
\gamma&=&\beta(\delta-1),  \label{Scaling} \\
2-\eta&=&\gamma/\nu  \ \ \ \ \ \ \ \ \ \ \ ({\rm Fisher's \ relation}), \label{Fisher} \\
\alpha+2\beta+\gamma&=&2 \ \ \ \ \ \ \ \ \ \ \ \ \ \ ({\rm Rushbrooke's \ relation}), \label{Rushbrooke} \\
2-\alpha&=&(d+z)\nu  \ \ \ \ \ ({\rm Josephson's \ relation}). \label{Josephson}
\end{eqnarray}
All the scaling laws here can be derived from Eq.~(\ref{Ehyperscaling}).%~\cite{PhysRevB.75.115121}.

Since the metal is characterized by a nonzero carrier density $X$ as the natural order parameter 
in distinction from the insulator ($X=0$), the unique length scale $\xi$ that diverges at the MQCP must be the mean carrier distance given by 
\begin{equation}
\xi\propto X^{1/d}.
\label{xi_X}
\end{equation}
In this case, we obtain
\begin{eqnarray}
\label{scaling_relation2}
\delta=z/d.
\end{eqnarray} 
%where $d$ is the spatial dimension.
The relation holds for both the bandwidth- and filling-control transitions. In the bandwidth-control case, $X$ in the metallic phase is the density of unbound doublon (double occupancy site) and holon (electron empty site). 
The last available scaling relation is 
\begin{eqnarray}
\label{scaling_relation3}
\nu=\beta/d.
\end{eqnarray}  
See Ref.~\onlinecite{PhysRevB.75.115121} and Methods C for the derivation of the scaling laws. 
%For the derivation of these relations see Ref.\onlinecite{MisawaImada2007} and for self-contained description, it is summarized in Supplementary Information. 

By using these relations, if only $\beta=q$ and $\nu z=p$ are known, other exponents can be obtained for $d=2$ as
$\alpha =2-(p+q), \gamma=p-q, \delta=p/q, \eta=4-2p/q, \nu=q/2$ and $z=2p/q$.
By using the values $p\sim 1.13\pm 0.19$ and $q\sim 0.97\pm 0.05$
%\txbs{$p\sim 1.0\pm 0.06$ and $q\sim 1.24\pm 0.18$}  
obtained by our simulation, we find the exponents listed in Fig.~\ref{fig:phase}{\bf b},
%$\alpha = -0.10\pm 0.24, \gamma = 0.16\pm 0.24, \delta= 1.16 \pm 0.20, \eta=1.68\pm 0.41, \nu=0.49\pm 0.02$ and $z= 2.32\pm0.41$,
%\txbs{$\alpha = -0.24\pm 0.24, \gamma = 0.24\pm 0.24, \delta= 1.24 \pm 0.24,$} 
%\txbs{$\eta=1.52\pm 0.48, \nu=0.5\pm 0.12$ and $z= 2.48\pm0.48$},
which can be consistent with 
$\alpha \sim 0, \gamma\sim 0, \delta\sim 1.0, \eta\sim 2.0, \nu\sim 1/2$ and $z\sim 2$.
In fact, our numerical result obtained independently from the scaling of $D-D_c$ indicates $\delta\sim 1$, which is consistent with this prediction. \txr{Furthermore, the spatial correlation of the double occupancy $D$ can be used to estimate $z+\eta$ independently from the above estimates, and though the estimate contains a large error bar, it suggests $z+\eta\sim 3.3 \pm 0.8$ (see  Sec.~\ref{Method} C and Sec. G of SM), which is again consistent with 4.0 estimated from the scaling theory.}  

\section{Discussion and Summary} \label{Discussion and Summary}
\vspace{5mm}
The quantum critical exponent $\nu z\sim 0.6\sim 0.9$ was indirectly estimated %at finite temperature 
above the classical Ising-type critical temperature %representing the endpoint 
of the first-order Mott transition, \minw{aiming at estimating the quantum criticality} by calculating the resistivity along the Widom line continued above the  critical temperature \mi{by using the DMFT~\cite{PhysRevLett.107.026401,PhysRevB.88.075143}. It \minw{was} compared with experimental measurements of organic solids, semiconductor moir\'{e} superlattices and transition metal dichalcogenides, because they all infer the $T=0$ criticality again from the Widom line~\cite{FurukawaKanoda2015,Moon2021,Li2021}. }
They also argued that the exponent does not appreciably change with the character of the neighboring phases~\cite{FurukawaKanoda2015} %(irrespective of magnetic and conducting properties) 
%antiferromagnetic or spin liquid in the insulator side and normal or superconducting phases in the metallic side, 
implying a universal and robust criticality.
Ambiguities of the definition of the Widom line \mi{and the estimate at temperature above nonzero critical temperature}, however, have yielded a variety of estimates for the exponent. By taking into account this ambiguity and also possible errors often recognized in the exponents estimated from the collapse to a single scaling plot employed by them (see also the next paragraph),
%unless the data really close to the critical point are available, 
and by considering a considerable variation of their estimates do not necessarily contradict our estimate of $\nu z\sim 1.13 \pm 0.19$. 

\minw{More importantly, the estimate by the DMFT~\cite{PhysRevLett.107.026401,PhysRevB.88.075143} is rigorous at infinite dimensions and the exponents can be different from the present two dimensional case.} 
\minw{Another DMFT study~\cite{PhysRevB.100.155152}  suggested that the estimated $\nu z$ in Ref.\onlinecite{PhysRevLett.107.026401} is related to the exponent of the instability line of the metastable insulating state at the boundary of the coexisting region. This instability line should vanish if the finite temperature critical temperature is lowered to zero as in the MQCP.  Therefore in this regard as well, $\nu z$ estimated along the Widom line may  not necessarily have a connection to the MQCP exponent studied here. If one wishes to estimate the MQCP exponents focused in this article, it is desired to estimate the exponent by sufficiently suppressing the critical temperature both in the theoretical and experimental studies. }
%\mi{It is not clear whether the inferrence of $T=0$ MQCP criticality from high temperatures above $T_{\rm c}$ captures only the crossover from the MQCP or it contains some mixtured influence from the quantum critical line in Fig.~\ref{MQCPSchematic}.}  
\mi{Our analysis %based on the quantitative estimates 
has determined a more comprehensive and quantitative set of various exponents $\beta, \delta, \nu z$ and $z+\eta$ from the scaling of four independent quantities including the double occupancy and charge gap, by straightforward estimates directly at zero temperature  precisely for the MQCP. The four exponents are shown to satisfy a perfect consistency with the scaling theory and determine all the exponents.}   

Though we obtained $d+z\sim 4$ as if it were at the upper critical dimension \txg{of the conventional symmetry-breaking magnetic transition}, it does not necessarily mean that the simple mean-field treatment is justified, because the Mott transition is not primarily a symmetry-breaking transition. Indeed, the anomalous dimension drives the nonzero and a fairly large exponent for $\eta$ ($\sim 2$), which can be analyzed as a Lifshitz-type topological transition that makes vanishing Fermi-surface pocket~\cite{Yamaji2006}. In fact, the exponents $\gamma\sim 0$, $z\sim 2$ and $\delta\sim 1$ look similar to a case of the 2D Lifshitz transition described by the emergence of electron and hole pockets~\cite{Yamaji2006}. 
%\txm{On the generality of the present model to study MQCP.}

The exponents $\alpha=\gamma \sim0$ and $\delta\sim 1$ indicate that the bandwidth-controlled MQCP does not drive divergent fluctuations in the charge channel, because the susceptibilities (the second derivatives of the energy with respect to $t_{\perp}$ and $U$) are not divergent at the MQCP. 
This is also indicated by nonsingular dependence of the energy as a function of the electron density at the MQCP as is shown in Fig.~S5 of SM. %\ref{E_MQCP} 
This absence is in contrast with the filling-controlled MQCP, where the divergent charge fluctuations and the %instability to the 
charge inhomogeneity are obtained as a common property~\cite{misawa2014superconductivity,PhysRevB.97.045138,PhysRevB.98.205132}.  The charge instability is also tightly linked with a strong effective attraction of the carriers~\cite{PhysRevB.75.115121,ImadaSuzuki2019}, which may be absent here.
%. The bandwidth control here does not show such an enhanced fluctuation for the metal near the MQCP. 
%caused by the unbinding of the doublon and holon. 
This is obviously a disadvantageous aspect for the promotion for  the superconductivity. 
Since the present simple model and its MQCP do not have any special aspect or unique symmetry, the universality class found here may be a standard one applicable widely to 2D MQCP.

On the other hand, the antiferromagnetic transition \txg{does not contradict} mean-field like normal divergent fluctuation with divergent susceptibility as is clarified in %the analysis of the magnetic transition in 
Sec. F of SM.  
The antiferromagnetic transition seems to occur at slightly larger $U$ ($U^{\rm AF}\sim 1.85$) than $U^{\rm MQCP}\sim 1.83$, but it is not easy to pin down whether they really differ (see SM F). 
%Nevertheless, the critical exponents characterized by $\nu^{\rm AF}\sim 0.5$ and $\eta^{\rm AF} +z^{\rm AF}\sim 2$ with resultant $\beta^{\rm AF}\sim0.5$ imply that the mean field values are satisfied because of $d+z^{\rm AF}\sim 4$ and $\eta^{\rm AF}\sim 0$, which implies that the system is above the upper critical dimension 
\txg{Nevertheless, the estimated $\nu^{\rm AF}\sim 0.5$ and $\eta^{\rm AF} +z^{\rm AF}\sim 2$ definitely indicate divergent fluctuations characterized by $\gamma^{\rm AF} >0$ and $\delta^{\rm AF} >1$ with the help of the scaling law independently of the Mott criticality.}
%\txr{We leave two possibilities for future study. The first is the simultaneous metal-insulator and antiferromagnetic transition at $U^{\rm MQCP}$, which is supported by our result, $\nu^{\rm MQCP}\sim \nu^{\rm AF}\sim 1/2$ and $z^{\rm MQCP}\sim z^{\rm AF}\sim 2$ implying that the scaling of the diverging spatial and temporal length that governs the criticality is the same. The other possibility is $U^{\rm AF}>U^{\rm MQCP}$, where the both diverging length scales are still governed by MQCP.}
\minw{
In any case, in the scaling properties, metal-insulator transition at the MQCP and the antiferromagnetic transition are decoupled as we show in Sec. I of SM. Therefore, the universality and critical exponents of the MQCP are not affected by either antiferromagnetic or paramagnetic nature of the insulating phase and the present system is expected to represent the general and universal band-width controlled 2D Mott transition. }

\minw{We also note that the spin and charge correlations show essentially 2D isotropic correlations as we see in Figs. S9 and S10 and manifests the 2D nature at the MQCP.}

\mi{We summarize the significance of the present paper:
\begin{itemize} 
\item[1.] The comprehensive set of critical exponents $\beta, \gamma, \delta, \eta, \nu$ and z, is estimated with consistency with the scaling theory. Our estimate provides us with a unified understanding of the universality class of clean D=2 MQCP for the bandwidth-controlled Mott transition. This is the same situation that the experiments in the literature aimed at. 
\item[2.] The exponents are estimated directly at $T=0$ unlike most of the previous studies.
\item[3.] The employed numerical method is a state-of-the-art quantum many-body solver provided as the open-source software mVMC, which can treat spatial and temporal quantum fluctuations.
\item[4.]  The present comprehensive clarification opens up a number of possible routes to test by experimental studies for complete  understanding of quantum Mott transition and nearby strange metal, which is expected to serve for future design of functionality.
\end{itemize}
}

\vspace{5mm}
%\noindent
\section{Methods} \label{Method}
%%%%%%%%%%%%%%%%%%%%%%%%%%%%%%%%%%%%%
%\noindent

\subsection{Numerical Method}\label{Numerical Method}
For the ground-state calculations, we employ a variational Monte Carlo (VMC) method~\cite{JPSJ.77.114701, PhysRevB.90.115137}.
The optimization procedure of the  VMC method to reach the ground state is equivalent to the imaginary time ($\tau)$ evolution represented by the repeated operation of $\exp(-\tau H)$ for the Hamiltonian $H$ or equivalently natural gradient method~\cite{PhysRevB.64.024512,JPSJ.85.034601}.  
We choose the periodic-antiperiodic boundary condition, i.e. $x (y)$-direction is periodic (antiperiodic) because its boundary condition allows closed shell condition for $L\times L = 4n\times 4n$ lattices, which makes the optimization of the variational parameters easier and statistical errors smaller due to the reduced degeneracy.  It also makes the extrapolation to the thermodynamic limit easier in the later procedure.
We use the trial wave function with correlation factors and the spin quantum-number projection as
\begin{align}
|\psi\rangle ={\mathcal L}^S{\mathcal P}_{\rm G} {\mathcal P}_{\rm J} {\mathcal P}^{(4)}_{\rm dh}|\phi_{\rm pair}\rangle,
\label{VMC_WF}
\end{align}  
where ${\mathcal P}_{\rm G},\; {\mathcal P}_{\rm J},\; {\mathcal P}^{(4)}_{\rm dh}$ are Gutzwiller, Jastrow and doublon-holon correlation factors and ${\mathcal L}^S$ is the spin quantum-number projection.
First, we give the pair-product wave function, defined as
\begin{align}
|\phi_{\rm pair}\rangle=\left(\sum^{N_s}_{i,j=1}f_{ij}\hat{c}^{\dag}_{i\uparrow}\hat{c}^{\dag}_{j\downarrow}\right)^{N_e/2}|0\rangle,
\label{Eq:PFF}
\end{align}
where $N_s$ is the number of sites and  $N_e$ is the number of electrons. 
%\txbs{
This wave function has the same form as the Bardeen-Cooper-Schrieffer (BCS) wave function, in which the spins are always restricted to pairs of up and down spins representing the singlet. \mi{The pair product function can also represent any form of the Slater determinant and in addition it has representability of any mean-field solution including magnetic, charge and superconducting symmetry breaking.}
%}

The averaged double occupancy 
\begin{equation}
D=\sum_{i}\langle \hat{n}_{i\uparrow}\hat{n}_{i\downarrow}\rangle /N_s,
\label{Double_Occupancy}
\end{equation}
where $\hat{n}_{i\uparrow}\;(\hat{n}_{i\downarrow})$ is the number operator of spin-up (spin-down) electrons, is a key quantity to understand strong correlation effects, especially in the Hubbard model, where $\langle \cdots \rangle=\langle \psi |\cdots | \psi \rangle/\langle \psi |\psi \rangle$ is the expectation value in the ground state.
In fact, the double occupancy is controlled by the Gutzwiller factor~\cite{Gutz}   
\begin{align}
  {\mathcal P}_{\rm G}=\exp\left(-g\sum_{i}\hat{n}_{i\uparrow}\hat{n}_{i\downarrow}\right)
\end{align}
to lower the energy where $g$ is a  variational parameter.

To take into account the long-ranged charge correlation, we also introduce the Jastrow factor~\cite{Jas} 
\begin{align}
{\mathcal P}_{\rm J}=\exp\left(-\frac{1}{2}\sum_{i\neq j}v_{ij}\hat{n}_i \hat{n}_j\right),
\end{align}
where $v_{ij}$ are variational parameters and $\hat{n}_i\equiv\hat{n}_{i\uparrow}+\hat{n}_{i\downarrow}$ is the number operator of electrons.
%\txbs{After the optimization, $v_{ij}$ directly reflect the charge-charge correlations.}

To express the correlation between doublon (site doubly occupied by the spin up and down electrons) and holon (empty site) in the strongly correlated regions,
we introduce a four-site doublon-holon correlation factor
\begin{align}
{\mathcal P}^{(4)}_{\rm dh}=\exp\left(-\sum^4_{m=0}\sum^{N_s}_{i=1}\left(\alpha^{\rm d}_{m}\xi^{\rm d}_{im}+\alpha^{\rm h}_{m}\xi^{\rm h}_{im}\right)\right),
\end{align}
where $\xi^{\rm d (h)}_{im}$ denotes the number operator of doublon (holon) around $i$th site. and $\alpha^{\rm d (h)}_{im}$ are the variational parameters.
We can express the operator $\xi^{\rm d (h)}_{im}$, for example, as
$\xi^{\rm d}_{i4}\equiv \hat{D}_i \prod_{\ell} \hat{H}_{i+\ell}$ 
%where $i+\ell$ is the nearest neighbor site of $i$
%\txbs{(this quantity is 1 only when all the nearest-neighbor sites are occupied by the holons)}
 and $\xi^{\rm h}_{i0}\equiv \hat{H}_i \prod_{\tau} (1-\hat{D}_{i+\tau})$,
%\txbs{(this becomes 1 only when no doublon is there at all the nearest-neighbor sites)},
where $i+\ell$ and $i+\tau$ run the nearest-neighbor sites around $i$ and $\hat{D}_i$ ($\hat{H}_i$) is the doublon (holon) operator defined as
$\hat{D}_i=\hat{n}_{i\uparrow}\hat{n}_{i\downarrow}$ ($\hat{H}_i=(1-\hat{n}_{i\uparrow})(1-\hat{n}_{i\downarrow})$).   

We set the $2\times 2$-sublattice structure for the pairing wave function $|\phi_{\rm pair}\rangle$ to reduce the variational parameters.
%In this chapter, we consider two initial trial states; one is the paramagnetic metal state that is the exact solution in the noninteracting case, and the other is the Hartree-Fock insulating solution assuming the antiferromagnetic (AF) order.

We calculate several physical quantities to identify the ground state. To determine the magnetic order and to distinguish a metal from an insulator,  we calculate relevant physical quantities, i.e. the momentum distribution function $n(\vec{k})$ and the spin structure factor $S(\vec{q})$.\\

Momentum distribution function $n(\vec{k})$ is given by
\begin{eqnarray}
n(\vec{k})=\frac{1}{2N_{s}}\sum_{i,j,\sigma}\left<\right.\hat{c}^{\dag}_{i\sigma}\hat{c}_{j\sigma}\left.\right>e^{i\vec{k}\cdot(\vec{r}_i-\vec{r}_j)},
\end{eqnarray}
where $\vec{r}_i$ is the vector representing the coordinate of $i$ th state.\\
In the same way, the spin structure factor $S(\vec{q})$ is calculated from
\begin{align}
S(\vec{q})&=\frac{1}{3N_{s}}\sum_{i,j}\left<\right.\hat{\vec{S}}_{i}\cdot\hat{\vec{S}}_{j}\left.\right>e^{i\vec{q}\cdot(\vec{r}_i-\vec{r}_j)}.
\end{align}

In the VMC calculations,
we prepared several different initial states (such as the paramagnetic metal (PM) (free fermion) and antiferromagnetic insulator (AFI) states)
%for the optimizations,
and optimized them until the variational parameters reach the convergence,
which may not necessarily preserve the character of the initial states and the nature of the optimized state is identified only after calculating physical quantities. 
To investigate the metal-insulator and magnetic transitions in the thermodynamic limit, 
we perform calculations of energy and other physical quantities
on the $L\times L$ site square lattice with the periodic-antiperiodic boundary condition for $L=16, 20, 24,$ and 28
% at
% $16\times 16$, $20\times 20$,  $24\times 24$ and  $28\times 28$ sites 
for each initial state
and the size dependences are examined.
%\txc{[$\ast$comment: $L$ was not defined]}

In this article, we perform the size extrapolations and scaling analyses
to examine the magnetic order and metallicity in the thermodynamic limit.
%based on optimized states regardless of those of initial states.

\mi{
This basic method is widely used and was tested from various perspectives in a number of benchmarks~\cite{PhysRevB.94.195126,PhysRevB.90.115137,Nomura2021}, ranging from 2D itinerant Hubbard model to frustrated quantum spin  models,
which has proven that it shows one of the best accuracy among available quantum many-body solvers with wide applicability to quantum lattice systems. In the present case, the ground state energy per site $E/N$ obtained from precisely the same VMC method using the form of the wave function Eq.(\ref{VMC_WF}) and the same Hamiltonian at the MQCP, $t_{\perp}=0.38$ and $U=1.83$ for $4\times 4$ lattice 
with the periodic-antiperiodic boundary condition is $-0.8665 \pm 0.0005$, while the value obtained from the exact diagonalization is -0.8700. The error $\sim$ 0.4\% is similar to the case of the benchmark in Ref. \onlinecite{PhysRevB.90.115137}. For physical quantities, the double 
occupancy $D=0.1869\pm0.0003$ and the peak of the spin structure factor $S(\vec{q})=0.4489\pm 0.0008$ at $(\pi,\pi)$ are compared 
with the exact values 0.1844 and 0.4301, respectively. This benchmark and that in the literature show that the accuracy well withstands and can be used for the 
present analyses. 
}
%%%%%%%%%%%%%%%%%%%%%%%%%%%%%%%%%%%%%%%%%%%%%%%%%%%%
%\appendix{\section{Definition of critical exponents}\label{Appendix_A_CriticalExponent}}
\subsection{Definition of critical exponents and derivation of scaling laws}\label{CriticalExponent}

%%%%%%%%%%%%%%%%%%%%%%%%%%%%%

Here, the double occupancy $D$
is regarded as a natural order parameter of the metal-insulator transition.
We calculate the critical exponents for the extrapolated double occupancy $D$ by controlling
$t_{\perp}$ and $U$, where the scheme for the extrapolation is given in SM G.
The exponent $\beta$ is defined from the asymptotic scaling form between the jumps of $D$ (namely, $\Delta D$) and 
$t_{\perp}$ measured from the critical point, i.e.
\begin{align}
\Delta D(t_{\perp})=a |t_{\perp}-t_{\perp}^{\rm MQCP}|^{\beta}
\label{beta}
\end{align}
near the MQCP point $t_{\perp}^{\rm MQCP}$, where $a$ is a constant. %\begin{align}
%\Delta D \propto |t_{\perp}-t^c_{\perp}|^{\beta},\label{beta}
%\end{align}
%and $\Delta D$ is the difference between $D_{\rm metal}$ and $D_{\rm ins}$ along the first-order transition line.
%\txbs{Here, $D_{\rm metal}$ and $D_{\rm ins}$ are the double occupation in the metallic and insulating phases at the critical point, respectively. 
%Equally, $\beta$ can also be defined as}
%\begin{align}
%\Delta D(U)=a' |U-U^{\rm MQCP}|^{\beta}
%\label{beta2}
%\end{align}
%\txbs{along the first-order transition boundary.}\txb{ [**Another definition of $\beta$  by $U$ is deleted.] }

The critical exponents $\delta$ and $\gamma$ are defined from 
\begin{align}
D-D^{\rm MQCP}|_{t_{\perp}=t^{\rm MQCP}_{\perp}} &\propto |U-U^{\rm MQCP}|^{1/\delta}, \label{delta}\\
\left.\frac{dD}{dU}\right|_{U=U_c}&\propto |t_{\perp}-t^{\rm MQCP}_{\perp}|^{-\gamma}. \label{gamma}
\end{align}  
The definition of the exponent $\alpha$ is given from

\begin{eqnarray}
\frac{d^2E}{dt^2_{\perp}}\propto |t_{\perp}-t^{\rm MQCP}_{\perp}|^{-\alpha}.\label{alpha}
\end{eqnarray}
for the ground-state energy $E$.

Insulators are distinguished from metals by a nonzero charge gap $\Delta_{\rm c}$, which is numerically defined by
\begin{align}
\label{Eq:gap}
\Delta_{\rm c} &\equiv \frac{1}{2}(\mu(N_{\uparrow}+N_{\downarrow}+1)-\mu(N_{\uparrow}+N_{\downarrow})),
\end{align}
 where the chemical potential $\mu$ is given as
$\mu(N_{\uparrow}+N_{\downarrow}+1)=(E(N_{\uparrow}+1,N_{\downarrow}+1)-E(N_{\uparrow},N_{\downarrow}))/2$, 
 and $E(N_{\uparrow},N_{\downarrow})$ is the optimized ground-state energy for systems 
 with the number of spin-up (spin-down) electrons $N_{\uparrow}$ ($N_{\downarrow}$).
The scaling of the charge gap around the MQCP at $U=U^{\rm MQCP}$ is defined as 
\begin{align}
\Delta_{\rm c}(U)=a_U |U-U^{\rm MQCP}|^{\nu z},
\label{gU}
\end{align}
where $\nu$ is the correlation-length exponent and $z$ is the dynamical exponent.
Here, $a_U$ is a constant.
This relation is the consequence of the scaling of the energy scale~\cite{PhysRevLett.107.026401,FurukawaKanoda2015},
$\Delta_{\rm c} \propto \xi^{-z}$, where $\xi$ is the unique length scale which diverges at the MQCP. The dynamical exponent relates the length (momentum)  to time (energy) scale and the correlation-length exponent $\nu$ is defined from 
\begin{equation}
\xi \propto |t_{\perp}-t_{\perp}^{\rm MQCP}|^{-\nu}.
\label{xi}
\end{equation}

Scaling relations Eqs.~(\ref{scaling_relation2}) and (\ref{scaling_relation3}) are derived in the following way~\cite{PhysRevB.75.115121}:
The scaling of the energy, Eq.~(\ref{Ehyperscaling}) is rewritten as $E\propto X^{(d+z)/d}$ by using Eq.(\ref{xi_X}).
By adding the $t_{\perp}$ and $U$ dependences, $E$ has the form
\begin{eqnarray}
E=-UX+B_0(t_{\perp}-t_{\perp}^{\rm MQCP})X^{\phi}+CX^{(d+z)/d}.
\end{eqnarray}
Minimizing $E$ for $t_{\perp}-t_{\perp}^{\rm MQCP}=0$ gives the scaling between $X$ and $U-U^{\rm MQCP}$, namely $\delta$ leading to Eq.(\ref{scaling_relation2}).  Eqs.(\ref{scaling_relation2}), (\ref{beta}) and (\ref{xi}) lead to 
%and for  $U=U^{\rm MQCP}$, the minimization of $E$ gives the scaling between $X$ and $t_{\perp}-t_{\perp}^{\rm MQCP}$ defined by $\beta$, yielding 
Eq.(\ref{scaling_relation3}).
%%%%%%%%%%%%%%%%%%%%%%%%%%%%%%%%%%%%%%%%%%%%%%%%%

\txr{The correlation of double occupancy is determined by}
\txb{\begin{eqnarray}
Q(\vec{r})=\frac{1}{N_s}\sum_{\vec{r}'} \langle (\hat{D}(\vec{r}+\vec{r}')-\langle \hat{D} \rangle)(\hat{D}(\vec{r}')-\langle \hat{D} \rangle)\rangle
\end{eqnarray}
where $\hat{D}(\vec{r})=\hat{n}_{\vec{r}\uparrow}\hat{n}_{\vec{r}\downarrow}$} \txr{ is the double occupancy operator
and } \txb{$\langle \hat{D} \rangle$} \txr{is the spatially averaged expectation value in the ground state. 
In the scaling hypothesis, this correlation is expected to follow}
\txb{
\begin{align}
Q(\vec{r}) \propto r^{-(d+z+\eta-2)}
\label{eq:D_corr}
\end{align}}
\txr{at asymptotically long distance $r=|\vec{r}|$.}
%\section{Methods for determination of metal-insulator and magnetic transitions}
%\subsection{\txr{M}etal-insulator transition}
%%%%%%%%%%%%%%%%%%%%%%%%%%%%%%%%%%%
\subsection{Methods for determination of metal-insulator transition and MQCP}
\label{MQCPchi2}
%%%%%%%%%%%%%%%%%%%%%%%%%%%%%%%%
In the region of first-order transitions, 
we see the energy level crossing between PM and AFI states, 
%as we can see in an example at $t_{\perp}=***$, 
which accompanies a jump of the double occupancy $\Delta D$.
The first-order transition point is identified by this energy level crossing after the system size extrapolation to the thermodynamic limit.
The metal\txb{-}insulator transition is corroborated by the opening of the charge gap and the qualitative change of the momentum distribution in Fig.~S4 in SM %\ref{nk_select} 
depicted for $t_{\perp}=0.5, 0.7$ and 1.0.
In most of the first-order region, we have confirmed that 
the transition indeed represents the simultaneous transition 
of metal-insulator and antiferromagnetic transitions
 by examining several relevant physical quantities around the transition point. %\txbs{(See Secs. C and F of SM)}.
%To determine the continuous transition point and its character, 
%we need some scaling plots with the size extrapolations. 
We have determined the continuous metal-insulator transition by the opening of the charge gap as is described in Fig.~S2 in SM  (See Sec. C  of SM). %\ref{fig:gap_all}. 

\begin{figure}[h!]
\begin{center}
\includegraphics[width=0.5\textwidth]{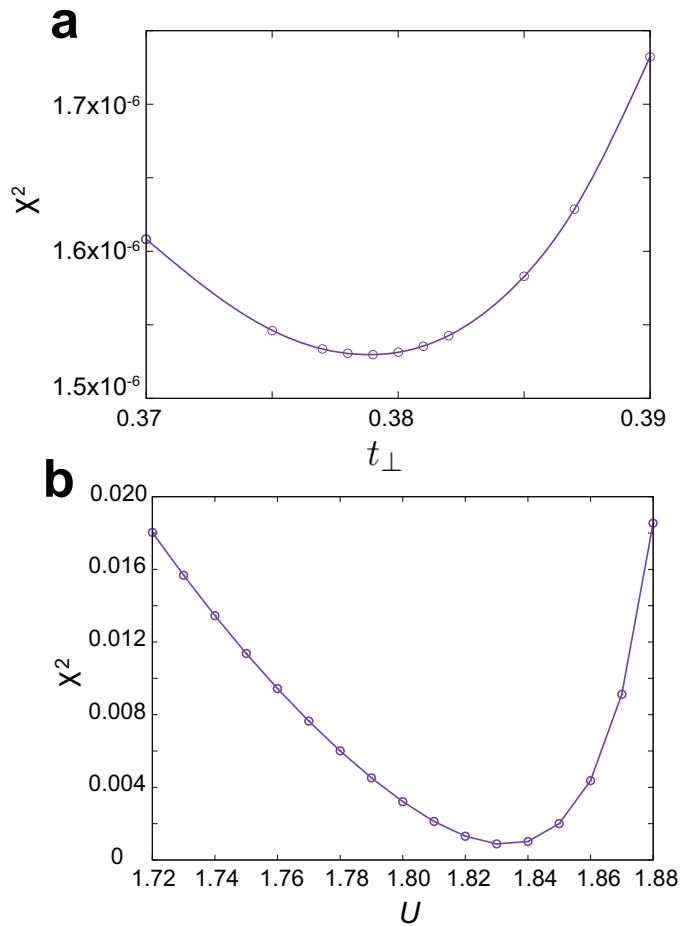}
\end{center}
\caption{
%\txb{[If possible, these figure will be shown with CDMFT's results.]}
{\bf a} $t_{\perp}$-dependence of
 $\chi^2$-value for the fitting Eq.(\ref{beta}), which results in $t_{\perp}^{\rm MQCP}=0.38\pm 0.05$ and $\beta=0.97\pm 0.05$.
{\bf b}  $U$ dependence of
 $\chi^2$-value for the fittings to combined Eq.(\ref{delta}), and  Eq.(\ref{gU}), which results in $U^{\rm MQCP}=1.83\pm 0.03$, $\nu z=1.13\pm0.19$,  $\delta_{\rm I}=0.98\pm0.03$ and $\delta_{\rm M}=1.05\pm0.04$.
%\txb{[***Two figures are merged.]}
}
\label{chi_tp-tpc}
\end{figure}

The MQCP point is first determined from the point where $\Delta D$ vanishes as is plotted in Fig.~\ref{fig:dD_exp}{\bf a}.
To determine $t_{\perp}^{\rm MQCP}$ and $\beta$ simultaneously,
%For this purpose, 
we have performed a regression analysis to optimize $t_{\perp}$ and $\beta$ dependences of $\Delta D$ in the form Eq.(\ref{beta})
by minimizing the following $\chi^2$,
%By assuming the $\chi^2$ distribution, $\chi^2$ value 
\begin{eqnarray}
\chi^2=\sum_i^{N_{t {\rm sample}}} ( \Delta D_i - \Delta D_{\rm fit})^2/(N_{t {\rm sample}}-2)
\label{chi2_t}
\end{eqnarray}
for $N_{t {\rm sample}}$ data point,
where $\Delta D_{\rm fit}$ has the form (\ref{beta}) and
$\Delta D_i$ is the simulation data. % at the same $t_{\perp}$.
The logarithmic difference is appropriate to
estimate the error for the power-law function.
In Fig.~\ref{chi_tp-tpc}{\bf b}, $t_{\perp}$ dependence of $\chi^2$ is plotted for the optimized exponent $\beta$.
From the minimum of $\chi^2$, $t_{\perp}^{\rm MQCP}$ is determined as $0.38 \pm 0.05$, where $\beta$ is $0.97 \pm 0.05$.
% as shown in Fig.~\ref{chi_tp-tpc}.  
The error bar is estimated from the bootstrap analysis explained in Methods E. %\end{document}

Since the MQCP can be signaled by the criticality given by the exponents $\beta, \delta$, $\nu z$, and the opening of the charge gap, the value of $U^{\rm MQCP}$ is estimated by the combined analysis of these three by employing $t_{\perp}^{\rm MQCP}=0.38$ as is analyzed in Fig.~\ref{chi_tp-tpc}{\bf a}, where the minimum of the $\chi^2$ value now defined as
%\txb{
\begin{eqnarray}
\chi^2&=&\sum_i^{N_{U {\rm sample}}}(\ln \Delta_{ci} -\ln \Delta_{c \rm fit})^2/(N_{U {\rm sample}}-2)\nonumber \\
&+&\sum_i^{N_{U {\rm I-sample}}} \frac{(\ln |D_{I i}- D^{\rm MQCP}| 
-\ln |D_{\rm fit}-D^{\rm MQCP}|)^2}{ 
N_{U {\rm I-sample}}-2}  \nonumber \\
&+&\sum_i^{N_{U {\rm M-sample}}} \frac{(\ln |D_{M i}- D^{\rm MQCP}| 
-\ln |D_{\rm fit}-D^{\rm MQCP}|)^2 }{ 
N_{U {\rm M-sample}}-2} \nonumber \\
\label{chi2_U}
\end{eqnarray}
%}
suggests $U^{\rm MQCP}=1.83 \pm 0.03$,
%. The error bar 0.03 is again obtained from the bootstrap.
%\txb{By fixing $U^{\rm MQCP}$ as 1.81, we simultaneously obtain 
$\nu z=1.13\pm0.19$,  $\delta_{\rm I}=0.98\pm0.03$ and $\delta_{\rm M}=1.05\pm0.04$.
For fittings to obtain these critical exponents, we assume Eq.(\ref{delta}) and  Eq.(\ref{gU}).

%%%%%%%%%%%%%%%%%%%%%%%%%%%%%%%%%%%%%%%%%%%%%%%%%%%%
%    Figure 5
%%%%%%%%%%%%%%%%%%%%%%%%%%%%%%%%%%%%%%%%%%%%%%%%%%%%

\subsection{Interpolation and bootstrap techniques}
To estimate %the error bars, we use the bootstrap method. For instance, to estimate the error bar of %mangetic and 
metal-insulator transition points, %with plausible error bars,
we introduce the interpolation techniques by fitting the computed data to an assumed form.
For reliable estimates for metal-insulator transition points, 
we interpolate energy and double occupancy data as a function of $U$ by the cubic function as
\begin{align}
\label{Eq:fitE}
f(U)=a_0 U^3+a_1 U^2+ a_2 U+a_3
\end{align} 
as the best fit of the $U$ dependence of quantities.
The crossing point of the interpolated energy of each metallic and insulating state gives us a reliable estimate of the level crossing point for the first-order transition.

In addition, we estimate the error bar of the level crossing point by using the bootstrap method.
Ground-state energy estimated by our Monte Carlo calculation, $E_{\rm MC}$ contains statistical errors 
%\txr{estimated by} independently \txr{performed} \txrs{determined by} Monte Carlo sampling 
given by the standard deviation $\sigma_{\rm MC}$.
Namely, we assume that $E_{\rm MC}$ obeys the Gaussian distribution $P(E_{\rm MC},\sigma^{2}_{\rm MC})$ and perform the following procedure:
%Especially for the energies, we estimate the errors as following.}
\begin{enumerate}
\item Generate a number of synthetic samples of the energy which follows the probability  $P(E_{\rm MC},\sigma^{2}_{\rm MC})$ around the interpolated $U$ dependence of the energy given by Eq.~(\ref{Eq:fitE}) for both insulating and metallic states.
%\item \txr{The interpolated energy curve is obtained by using Eq. (\ref{Eq:fitE}).} 
\item Calculate the crossing point between the insulating and metallic states for each synthetic data.
\item  Calculate the variance of the crossing points of the synthetic data, which gives the estimate of the error bar.
\end{enumerate} 
Furthermore, we also apply the bootstrap method for determining statistical errors for critical exponents and $t_{\perp}^{\rm MQCP}$ and $U^{\rm MQCP}$ in Methods D.

\vspace{5mm}
\begin{acknowledgments}
%%%%%%%%%%%%%%%%%%%%%%%%%%%%%%%%%%%%%%%%%%%%%%%%%%%%
\txr{The authors acknowledge Macin Raczkowski for useful discussions.
This work was supported in part by KAKENHI Grant No.16H06345 and 22A202 from JSPS. 
This research was also supported by MEXT as ``program for Promoting Researches on the Supercomputer Fugaku"(Basic Science for Emergence and Functionality in Quantum Matter - Innovative Strongly Correlated Electron Science by Integration of Fugaku and Frontier Experiments -, JPMXP1020200104). 
We thank the Supercomputer Center, the Institute for Solid State Physics, The University of Tokyo for the use of the facilities.
We also thank the computational resources of supercomputer Fugaku provided by the RIKEN Center for Computational Science (Project ID: hp210163, hp220166) and Oakbridge-CX in the Information Technology Center, The University of Tokyo.} \txg{FFA thanks the DFG  for  funding via the  Wurzburg-Dresden Cluster of Excellence on Complexity and Topology in Quantum Matter ct.qmat (EXC 2147, project-id 390858490).}\\
\end{acknowledgments}

%\bibliographystyle{apsrev}% Produces the bibliography via BibTeX.
%\bibliography{masterbib}
%Merlin.mbs v4.21 2009-07-09.
%

\newpage

\noindent
{\Large \bf Supplementary Material} \\
\renewcommand{\thefigure}{S\arabic{figure}}
\setcounter{figure}{0}
\renewcommand{\theequation}{S\arabic{equation}}
\setcounter{equation}{0}
%%%%%%%%%%%%%%%%%%%%%%%%%%%%%%%%%%%%%%%%%%%%%%%%%%%%
%\subsection{Shape of Fermi surface for noninteracting system}

\noindent
{\bf A. Shape of Fermi surface for noninteracting system} \\
%%%%%%%%%%%%%%%%%%%%%%%%%%%%%%%%%%%%%%%%%%%%%%%%%%%%
%    Figure S1
%%%%%%%%%%%%%%%%%%%%%%%%%%%%%%%%%%%%%%%%%%%%%%%%%%%%
\noindent
\begin{figure}[h!]
\begin{center}
\includegraphics[width=8cm]{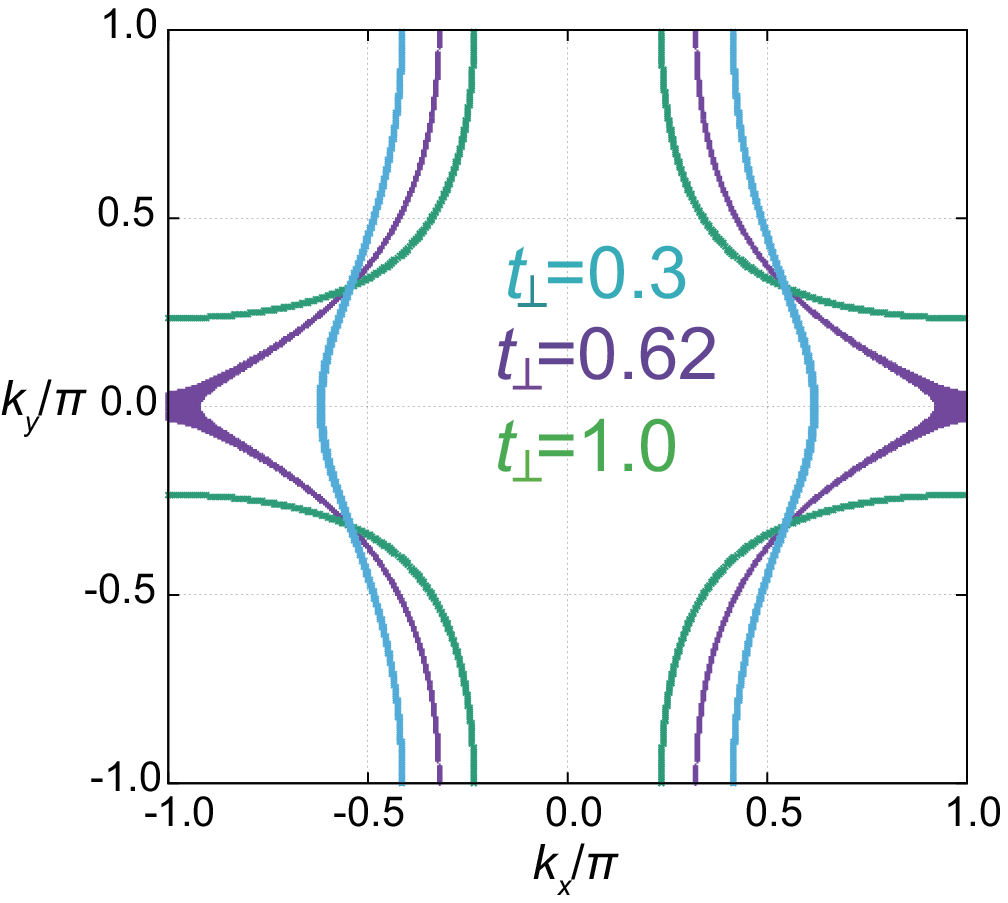}
\end{center}
\caption{\label{fig:fs2}
Fermi surfaces of the model for $U=0$ at half filling. 
 Green, purple and light blue curves indicate the Fermi surface at $t_{\perp}=1, 0.62$ and 0.3, respectively.}
\end{figure}
%%%%%%%%%%%%%%%%%%%%%%%%%%%%%%%%%%%%%%%%%%%%%%%%%%%%
Figure~\ref{fig:fs2} shows the Fermi surface for $U=0$.

\begin{figure*}[tbh]
\begin{center}
\includegraphics[width=12cm]{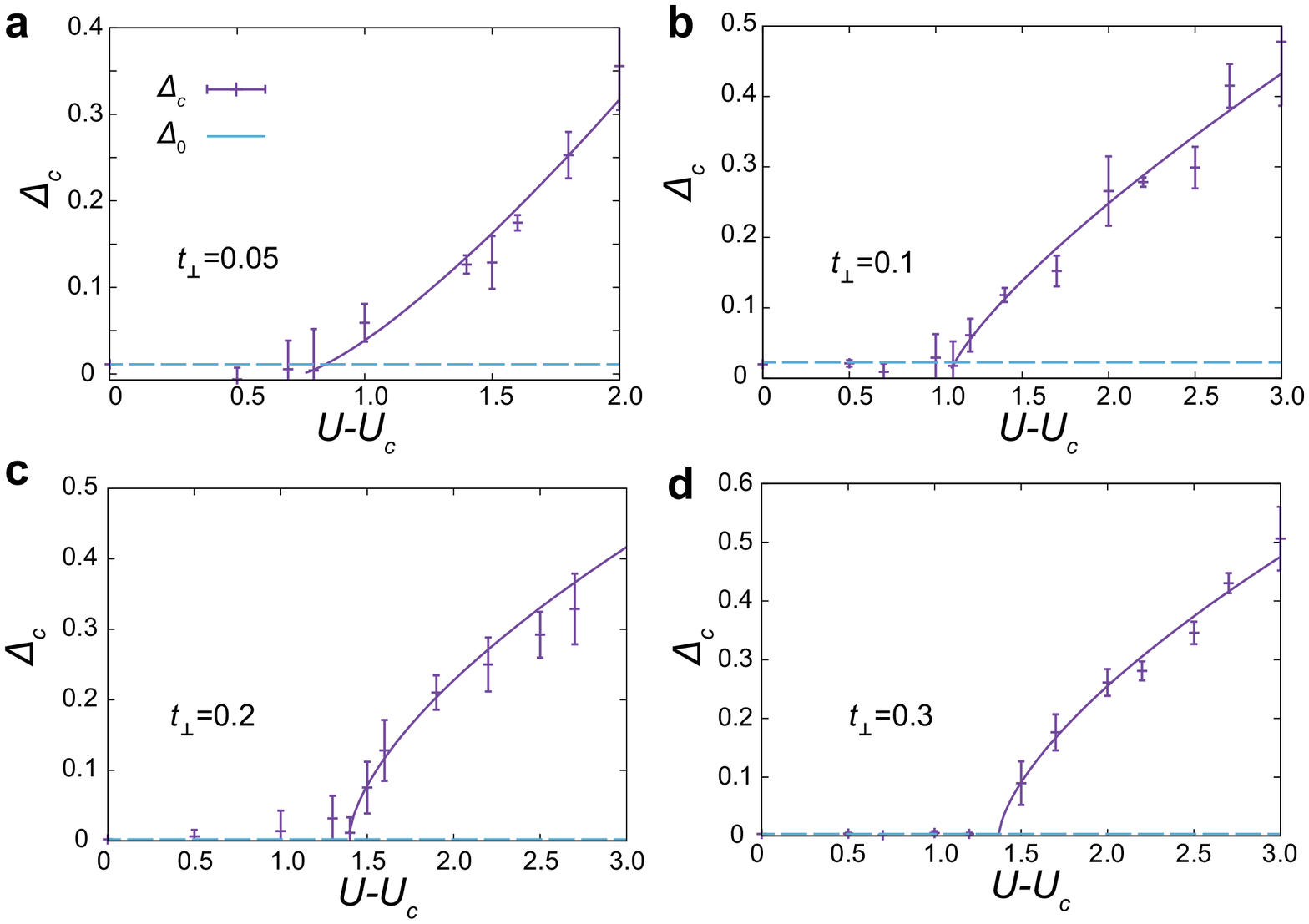}
\end{center}
\caption{\label{fig:gap_all}
Examples of charge gaps $\Delta_{\rm c}$ for %$20\times 20$, $24\times 24$ and 
$28\times 28$ sites in the cases of  
$t_{\perp}=0.05$({\bf a.}), 0.1({\bf b.}), 0.2({\bf c.}) and 0.3({\bf d.}). Thin dashed blue lines are the finite-size gaps in the noninteracting case.}
%\end{center}
\end{figure*}

%%%%%%%%%%%%%%%%%%%%%%%%%%%%%%%%%%%%%%%%%%%%%%%%%%%%%%
%\subsection{Charge gaps in continous transition region}
\vskip 5mm
\noindent
{\bf B.  Charge gaps in continous transition region} \\
%\txrs{If the energy-level crossing is clear, we are able to judge the bounday of the metal-insulator transition. However, it is difficult to judge it in the continuous transition region.}
%We investigate the chrge gaps.
Charge gap defined by Eq. (23) %Eq.(\ref{Eq:gap}) 
is obtained from the procedure in Methods C. An example of the calculated results for %$20\times 20$, $24\times 24$ and 
$28\times 28$ lattices  in the cases of
a. $ t_{\perp}=0.05$, b. 0.1, c. 0.2 and d. 0.3 are shown in Fig. \ref{fig:gap_all}. 
The phase boundary of metal-insulator transitions in Fig.~1 %\ref{fig:phase} 
is determined 
by analyzing these results. The size of the artifact by the finite-size gap $\Delta_0$ speculated from the noninteracting case is illustrated by the horizontal dotted line.  
%\txb{[Prof.Imada's comments: To increase the reliability, calculate the gap with smaller interval, say with the interval of delta U=0.1 near the gap closing and less than 0.2 for other regions.$\rightarrow$ Now, I'm trying to refine data especially for Figs. (c) and (e).]}
%%%%%%%%%%%%%%%%%%%%%%%%%%%%%%%%%%%%%%%%%%%%%%%%%%%%
%    Figure S2
%%%%%%%%%%%%%%%%%%%%%%%%%%%%%%%%%%%%%%%%%%%%%%%%%%%%

\begin{figure*}[htb]
\begin{center}
\includegraphics[width=16cm]{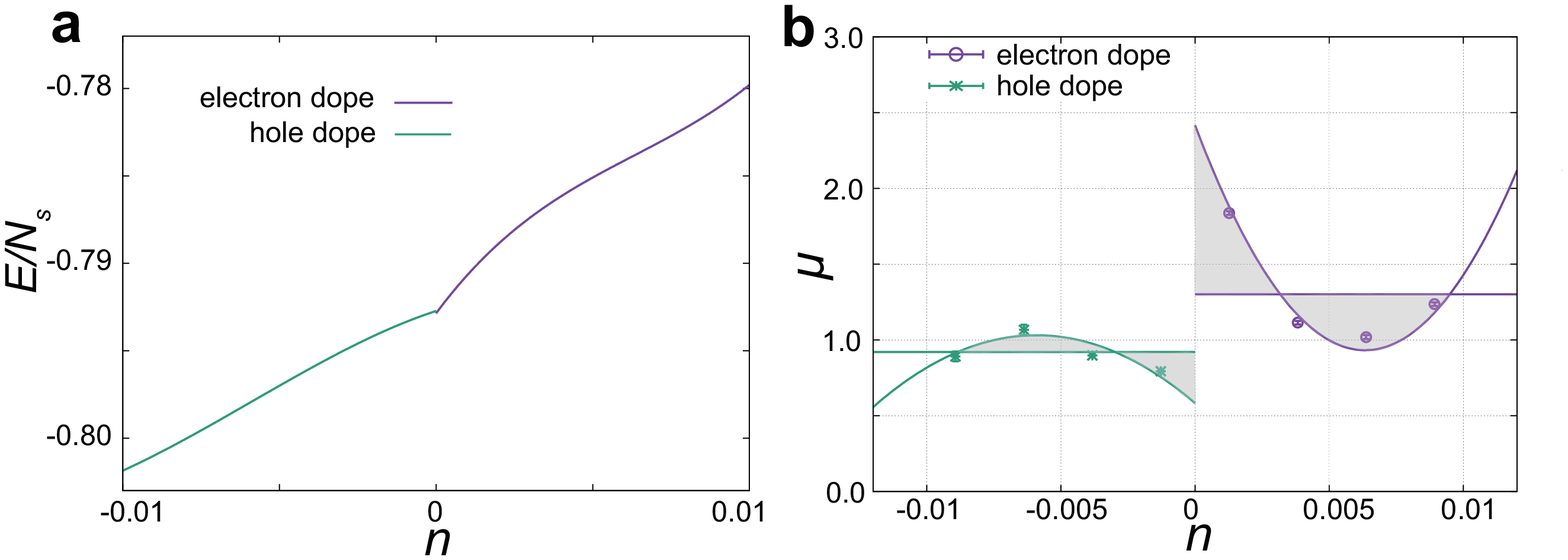}
\end{center}
\caption{\label{fig:chemipo}
An example of charge gap estimate by Maxwell rule. {\bf a.} Ground-state energy as a function of carrier density $n$ at $t_{\perp}=t^{\rm MQCP}_{\perp}=0.38$ and $U=2.5$ for $28\times 28$-site system.  {\bf b.} Chemical potential $\mu$ vs. $n$ for the same case as {\bf a.} The Horizontal solid lines in both electron- and hole-doped regions represent the lines used for the Maxwell's construction, where the horizontal line is drawn so as to make the areas of the two shaded regions surrounded by the line and the curve equal. For the curve and line adjacent to $n=0$, we count the area of the shaded domain surrounded by the curve, the horizontal line and the $n=0$ vertical line. Then the two crossing points of the line and the curve determines the two points of the phase separation for each electron and hole side as assured by the thermodynamic stability.
The difference in $\mu$ between electron and hole doped sides (the  difference of $\mu$ between the two horizontal lines) gives the charge gap.}%\txb{[**Fig.a. is updated.]}}
%\end{center}
\end{figure*}

\vskip 5mm
%%%%%%%%%%%%%%%%%%%%%%%%%%%%%%%%%%%%%%%%%%%%%%%%%%%%
%\subsection{Method to estimate of the charge gap in phase separation region}
\noindent
{\bf C.  Method to estimate the charge gap in phase separation region} \\
%%%%%%%%%%%%%%%%%%%%%%%%%%%%%%%%%%%%%%%%%%%%%%%%%%%%
Figure~\ref{fig:chemipo}.{\bf a.} shows that the energy as functions of $n$ shows phase separation for $U \geq 1.9$ in case of $t_{\perp}=0.38$. The carrier density $n$ is defined as $n=N_e/N_s-1/2$.
Figure~\ref{fig:chemipo}.{\bf b.}  illustrates the procedure to estimate the charge gap when the phase separation takes place. Convex (concave) downward curve of the chemical potential in the electron (hole) doped region indicate the phase separation. By drawing the horizontal line to make the area of the two regions surrounded by the horizontal line and the chemical potential curve, the phase separation region can be obtained, where the pinned chemical potential during the phase separation is given by the horizontal line.  
The difference in the pinned chemical potential between the  hole and electron doped sides is the charge gap.
%%%%%%%%%%%%%%%%%%%%%%%%%%%%%%%%%%%%%%%%%%%%%%%%%%%%
%    Figure S3
%%%%%%%%%%%%%%%%%%%%%%%%%%%%%%%%%%%%%%%%%%%%%%%%%%%%
%%%%%%%%%%%%%%%%%%%%%%%%%%%%%%%%%%%%%%%%%%%%%%%%%%%%

\begin{figure*}[htb]
\begin{center}
\includegraphics[width=11cm]{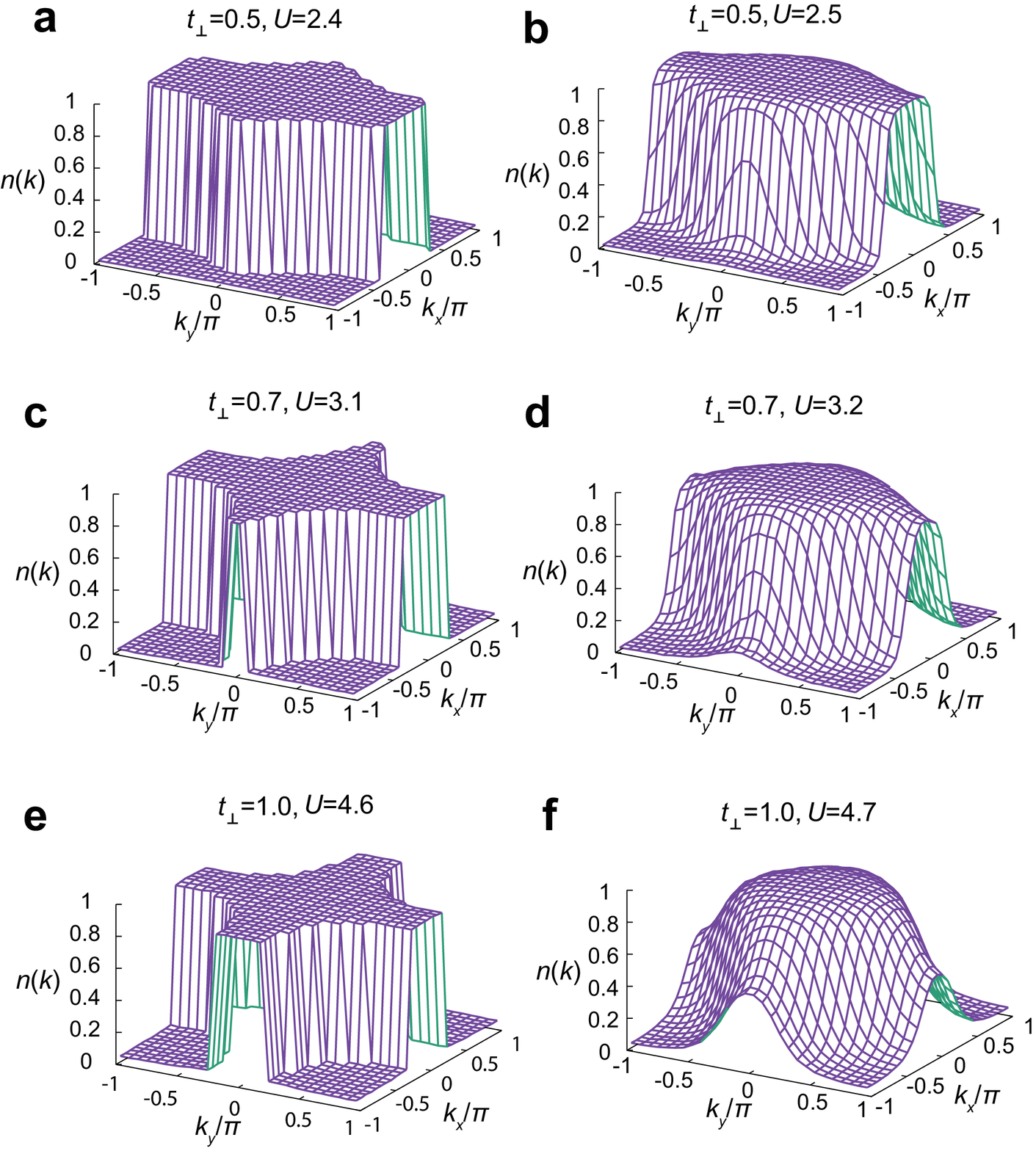}
\end{center}
\caption{\label{nk_select}
Momentum distribution functions $n(\vec{k})$ for $t_{\perp}=$0.5, 0.7 and 1.0 around the metal-insulator transition points. {\bf a., c., e.} indicate the metallic phase with sharp Fermi surface while {\bf b., d., f.} indicate the insulating phase without the Fermi surface.}
\end{figure*}

\vskip 5mm
%\subsection{Momentum distribution functions in the first-order transition region}
\noindent
{\bf D. Momentum distribution functions in first-order transition region} \\
In the most part of the first-order transition region, the metal-insulator and antiferromagnetic transition occur simultaneously. 
To confirm the metal-insulator transition, momentum distribution functions $n(\vec{k})$ around the energy level crossing points are shown in Fig.\ref{nk_select}, where
the shape of $n(\vec{k})$ is qualitatively different between the metal and the insulator.
%%%%%%%%%%%%%%%%%%%%%%%%%%%%%%%%%%%%%%%%%%%%%%%%%%%%
%    Figure S4
%%%%%%%%%%%%%%%%%%%%%%%%%%%%%%%%%%%%%%%%%%%%%%%%%%%%

\begin{figure*}[htb]
\begin{center}
\includegraphics[width=14cm]{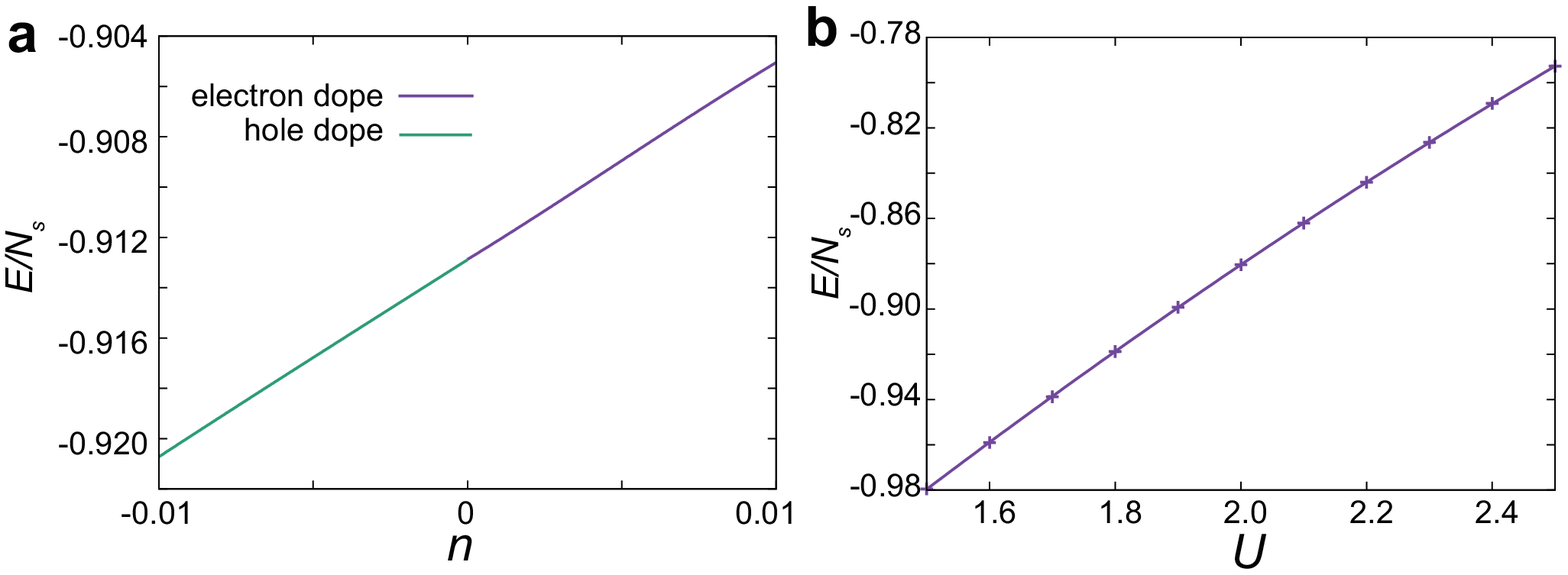}
\end{center}
\caption{\label{E_MQCP}
Ground-state energy as a function of carrier density $n$ ({\bf a.}) and $U$ ({\bf b.}), which support the absence of the singular behavior.}
%\end{center}
\end{figure*}

%\subsection{Absence of singularity in energy as function of control parameter}
\vskip 5mm
\noindent
{\bf E. Absence of singularity in energy as function of control parameter} \\
Figure~\ref{E_MQCP} shows that the energy as functions of $n$ ({\bf a.}) and $U$  ({\bf b.})  look nonsingular around the MQCP at $U=1.83$ and $t_{\perp}=0.38$. 
%%%%%%%%%%%%%%%%%%%%%%%%%%%%%%%%%%%%%%%%%%%%%%%%%%%%
%    Figure S5
%%%%%%%%%%%%%%%%%%%%%%%%%%%%%%%%%%%%%%%%%%%%%%%%%%%%
%%%%%%%%%%%%%%%%%%%%%%%%%%%%%%%%%%%%%%%%%%%%%%%%%%%%

%%%%%%%%%%%%%%%%%%%%%%%%%%%%%%%%
%     AF transition
%%%%%%%%%%%%%%%%%%%%%%%%%%%%%%
%%%%%%%%%%%%%%%%%%%%%%%%%%%%%%%%%%%%%%%%%%%%
%\subsection{Determination of antiferromagnetic transition and criticality}
\vskip 5mm
\noindent
{\bf F. Determination of antiferromagnetic transition and its criticality} \\
%%%%%%%%%%%%%%%%%%%%%%%%%%%%%%%%%%%%%%%%%%
%Since the antiferromagnetic transition takes place simultaneously with the transition to the insulator at the MQCP, it is important to clarify the relation of them.  
The universality of the magnetic transition may belong to a class different from the Mott transition.
As well as metal-insulator transitions, 
we see the clear jumps of the staggerred magnetization $m_s$ in the region of first-order transitions.
However,  the border between paramagnetic and antiferromagnetic phases is 
not straightforward in the region of continuous transitions.   
Here we first describe how $U^{\rm AF}$ is estimated and then the estimate of the critical exponent at the MQCP later. 

\subsubsection{Antiferromagnetic transition determined by correlation ratio method}
We determine the boundary of the antiferromagnetic phase by using
the correlation-ratio method~\cite{Kaul}, 
%and independently from the extrapolation of $S(\pi,\pi)/N_s$ 
where the correlation-ratio parameter $S_g$ obtained from the spin structure factor $S(q)$ is given by
\begin{align}
\label{Sg}
S_g\equiv 1-\frac{S(\pi,\pi+\Delta q_y)}{S(\pi,\pi)}.
\end{align}
Here, $\pi+\Delta q_y$ is the nearest-neighbor $\vec{k}$-point to $(\pi,\pi)$.
We plotted this ratio for $20\times 20$, $24\times 24$ and $28\times 28$ sites, to determine the border 
of paramagnetic and antiferromagnetic phases.
In the nonmagnetic region,  $S_g$ converges to zero with increasing system size, because $S(\vec{q})$ is finite and continuous in the thermodynamic limit.
On the other hand, in the AF region,  $S_g$ converges to one by increasing the system size.
It is empirically observed that the different-size curves cross and the crossing point does not sensitively depend on the system sizes, which serves as a good estimate of the transition point in thermodynamic limit~\cite{Kaul,Kaul2}.  
We plot the curves and their crossing points for $20\times 20$, $24\times 24$ and $28\times 28$ sites.

\begin{figure*}[htb]
\begin{center}
\includegraphics[width=14cm]{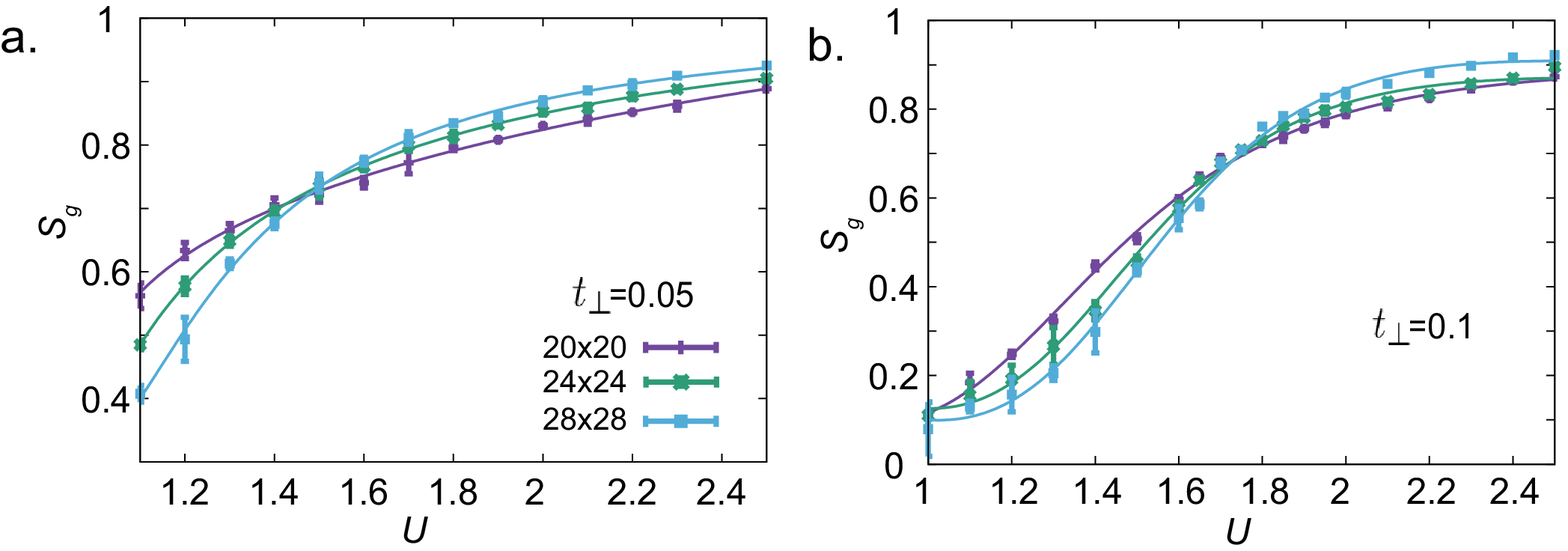}
\end{center}
\caption{\label{fig:CorrRatiotp0.05_0.1}
Correlation-ratio plot for $20\times 20$, $24\times 24$ and $28\times 28$ sites in the cases of  {\bf a.} $ t_{\perp}=0.05$ and {\bf b.} $ t_{\perp}=0.1$. The crossing points suggest $U^{\rm AF}\sim 1.48$ and 1.73, respectively.}
%\end{center}
\end{figure*}

\begin{figure*}[h!]
\begin{center}
\includegraphics[width=16cm]{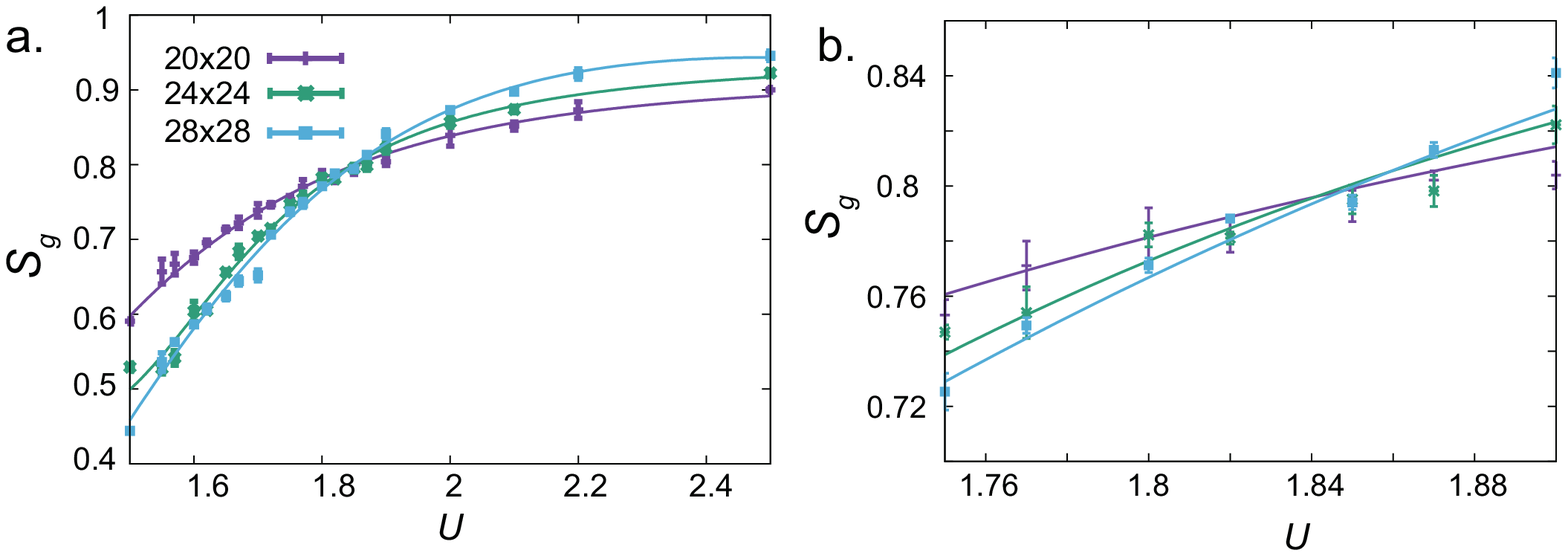}
\end{center}
\caption{\label{fig:sgms}
Determination of the antiferromagnetic transition along $t_{\perp}=t_{\perp}^{\rm MQCP}=0.38$.
Analysis by correlation-ratio method by using $20\times20$, $24\times24$ and $28\times28$ lattices is shown in wide region in {\bf a.} and the zoom-in plot in  {\bf b.}. 
%{\bf c.} Size extrapolations of $S(\pi,\pi)/N_s$ by $L^{-1}$ for 16x16, 20x20, 24x24 and 28x28 lattices.
}
\end{figure*}

In the same way as fittings of energy and double occupancy, we interpolate the correlation-ratio parameter $S_g$ as a function of $U$ by assuming the rational function as
\begin{align}
\label{Eq:fitSg}
g(U)=\frac{a_0 U^2+a_1 U+ a_2}{a_3 U^2+a_4 U+ a_5}.
\end{align}
From this fitting we are able to estimate the correlation-ratio crossing point by the interpolation. %\appendix{\section{Correlation-ratio parameter in the continous transition region}}
%\subsection{Correlation-ratio parameter in the continous transition region} \label{Correlation-ratio} In the continuous transition region, \txrs{we introduce} the correlaton-ratio parameter $S_g$ as Eq.(\ref{Sg}) \txr{is used}.
The phase boundary of the magnetic transition   
in Fig.~1 %\ref{fig:phase} 
is thus determined from  the crossing points of $S_g$
for $20\times 20$, $24\times 24$ and $28\times 28$ sites.
We show the correlation-ratio plot for 
$20\times 20$, $24\times 24$ and $28\times 28$ sites in the cases of $t_{\perp}=0.05$ and 0.1 in Fig. \ref{fig:CorrRatiotp0.05_0.1}, where the metal-insulator transition is clearly different and the quantum spin liquid phase (NMI) is found.  For $t_{\perp}=t_{\perp}^{\rm MQCP}=0.38$ the plot is shown in Fig.~\ref{fig:sgms}.
Then the magnetic transition point is consistently estimated as $U^{\rm AF}\sim1.85\pm 0.02$, which is close to $U^{\rm MQCP}\sim 1.83 \pm 0.03$, but seems to be slightly larger within the error bar. 
% the MQCP within the statistical accuracy indicating that the Mott and antiferromagnetic transition occurs simultaneously at the MQCP.
%%%%%%%%%%%%%%%%%%%%%%%%%%%%%%%%%%%%%%%%%%%%%%%%%%%%
%    Figure S6
%%%%%%%%%%%%%%%%%%%%%%%%%%%%%%%%%%%%%%%%%%%%%%%%%%%%
%%%%%%%%%%%%%%%%%%%%%%%%%%%%%%%%%%%%%%%%%%%%%%%%%%%%
%    Figure S7
%%%%%%%%%%%%%%%%%%%%%%%%%%%%%%%%%%%%%%%%%%%%%%%%%%%%

%%%%%%%%%%%%%%%%%%%%%%%%%%%%%%%%%%%%%%%%%%%
\subsubsection{Critical exponent at antiferromagnetic transition}
%%%%%%%%%%%%%%%%%%%%%%%%%%%%%%%%%%%%%%%%
%\txc{[$\ast$ what is $L$?]}
We here estimate the critical exponents for the antiferromagentic transition at the MQCP.
For this purpose, we adopt the finite-size scaling relation for the spin structure factor $S(q)$,
\begin{align}
\label{eq_sc}
S(\pi,\pi)=L^{-z+2-\eta} f_m(uL^{1/\nu}),
\end{align}
where $u=(U-U^{\rm AF})/U^{\rm AF}$ and $z$ represents the dynamical exponent while $f_m$ is a scaling function and $\eta$ is the exponent associated with the anomalous dimension.

%%%%%%%%%%%%%%%%%%%%%%%%%%%%%%%%%%%%%%%%%%%%%%%%%%%%
%    Figure S8
%%%%%%%%%%%%%%%%%%%%%%%%%%%%%%%%%%%%%%%%%%%%%%%%%%%%
\begin{figure}[h!]
\begin{center}
\includegraphics[width=8cm]{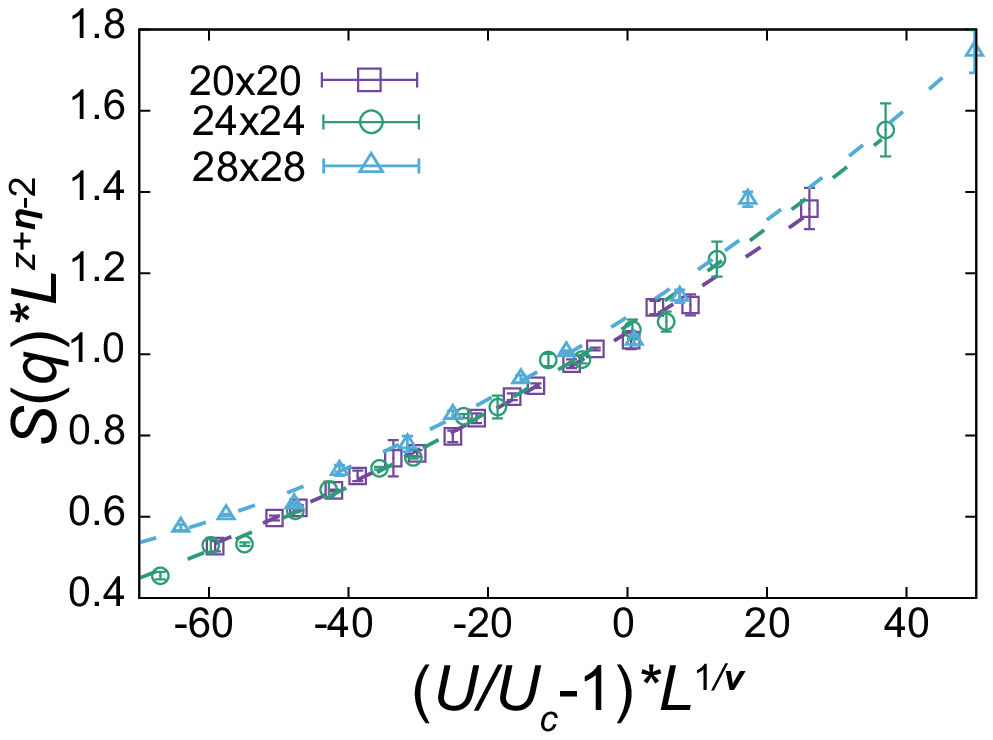}
\end{center}
\caption{\label{fig:scaling_sq}
Finite-size scaling of $S(\pi,\pi)$ by using scaling relation (\ref{eq_sc}).
The data collapse to a single scaling curve is found for $\nu=0.52\pm 0.02$, and $\eta+z=1.9\pm 0.1$ with $U^{\rm AF}=1.85 \pm 0.02$. 
}
\end{figure}
%%%%%%%%%%%%%%%%%%%%%%%%%%%%%%%%%%%%%%%%%%%%%%%%%%%%

\begin{figure}[h!]
\begin{center}
\includegraphics[width=0.6\textwidth]{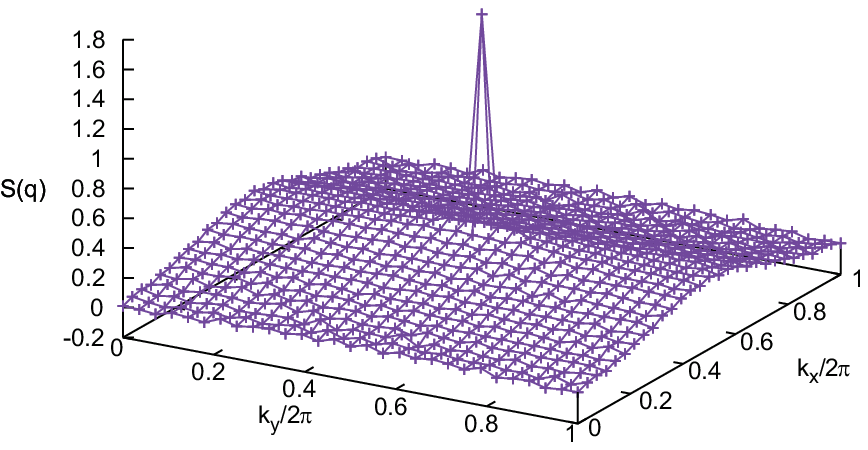}
\end{center}
\caption{\label{fig:S(q)}
Spin structure factor at MQCP.
}
\end{figure}

As shown in Fig.\ref{fig:scaling_sq}, we obtain the exponent as 
\begin{align}
\nu=0.52\pm 0.02, \ {\rm and} \  \eta+z=1.9\pm 0.1
\end{align}
if the scaling form (\ref{eq_sc}) is used with $U^{\rm AF}=1.85$ at $t_{\perp}=0.38$. 
%($U^{\rm AF}$ is the magnetic transition point for $t_{\perp}=t_{\perp}^{\rm MQCP}$ and is obtained by using the correlation ratio method from the crossing point of $S_g$ in Fig. \ref{fig:tp0.38}(e)).}
%We find $U^{c}$ coincides with $U_{\rm AF}$. 
Moreover, if we assume the hyperscaling relation
\begin{align}
\frac{\beta}{\nu}=\frac{z+d-2+\eta}{2},
\end{align}
we can estimate the critical exponent $\beta_{\rm AF}=0.49\pm 0.03$, 
%in the side of nonmaganetic insulater.
which turns out to be consistent with that of the mean-field theory ($\beta=0.5$). This is justified when $z\sim 2$ so that $d+z=4$ assures that the present system is located \txg{just at the upper critical dimension in the conventional framework of Ising or Hertz-Moriya~\cite{PhysRevB.14.1165,Moriya} and the critical exponents are marginally given by the mean-field values for the symmetry breaking transition. 
 This is also consistent with $z+\eta\sim 2$ resulting in $\eta\sim 0$, indicating the absence of the anomalous dimension. Then $\gamma=1$ and $\delta=3$ derived from the scaling relation} %\txb{[$\delta=1$? Please confirm the estimate.]}
indicate divergent fluctuations in contrast with the universality of the metal-insulator transition.
%with gapped particle-hole excitations). 
A large $z (\sim 2)$ instead of the normal value $z=1$ expected for the antiferromagnetic spin wave dispersion could be the consequence of the proximity from the MQCP. \txg{Instead, it is conceivable that non-negligible $\eta>0$ makes $d+z<4$ so that the deviation from the mean-field value exists, which may drive $z$ to decrease from 2, though the presence of the diverging fluctuations characterized by  $\gamma>1$ and $\delta>1$ would not change.  These issues} should be carefully examined in the future in the region close to the transition point if $U^{\rm AF}$ is different from $U^{\rm MQCP}$. \mi{Of course, the AF long-range order requires the multi-dimensionality $d\geq 2$ of the system. Although the background broad peak reflects the moderate anisotropy of the Hamiltonian, the spin structure factor $S(q)$ shown in Fig.~\ref{fig:S(q)} clearly demonstrates that the spin correlation is 2D isotropic behavior for a critical sharp peak at $(\pi,\pi)$ even at MQCP.}

\begin{figure}[htb]
\begin{center}
\includegraphics[width=0.6\textwidth]{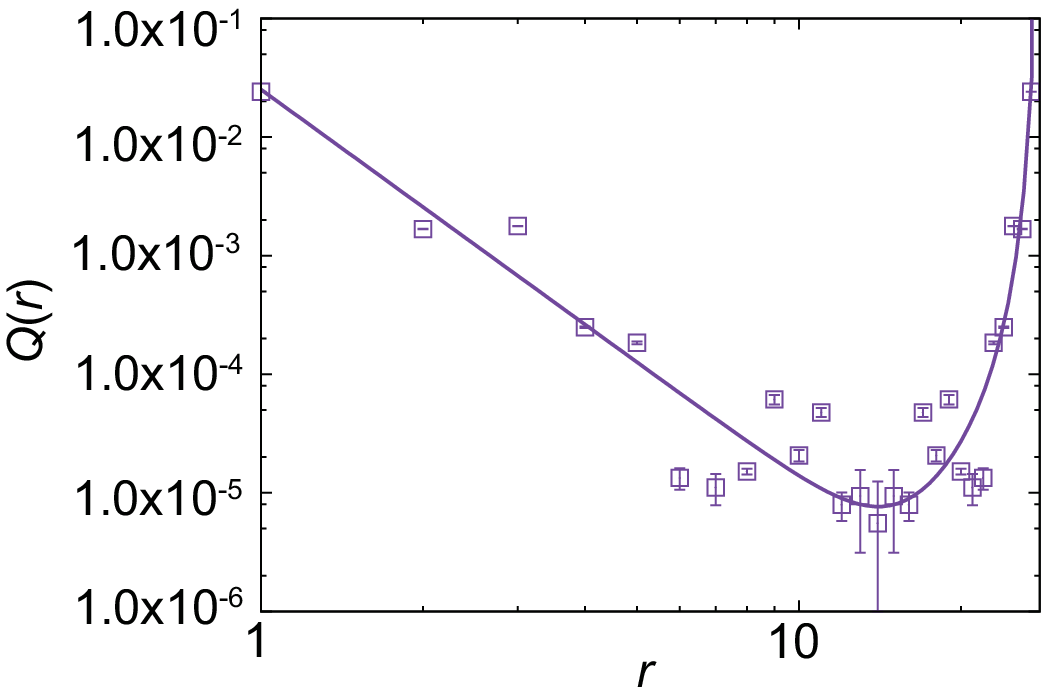}
\end{center}
\caption{\label{fig:D_corr}
\txr{
Spatial correlation of double-occupancy fluctuation in $x$ direction defined in Eq.~(26). Error bars are those of statistical errors in the Monte Carlo sampling. The fitting curve is obtained by using the form}
\txb{$Q(\vec{r}) \propto 1/|r|^{z+\eta}+\sum_{n\ne (0,0)}[(1/|r+L\vec{n}|^{z+\eta})-(1/|L\vec{n}|^{z+\eta})]$} \txr{ to fit the data at $r\ge 3$ to estimate the asymptotic form at long distance. It suggests more or less isotropic power-law decay with $z+\eta =3.3 \pm 0.8$ by taking account of  the combined errors of the fitting and the Monte Carlo sampling.
}
}
\end{figure}

\vskip 5mm
\noindent
{\bf G. Double occupancy correlation} \\
\txr{Spatial correlation of double occupancy $D$ is defined in Eq.~(12). }
\txr{The spatial correlation of the fluctuation of $D$ defined by Eq.~(26) is plotted in Figure~\ref{fig:D_corr}, where the fitting of} \txb{$Q(\vec{r})$ suggests $z+\eta=3.3 \pm 0.8$} \txr{from Eq.(27).
The value is consistent with the present scaling theory that requires $z+\eta=4$.}

\begin{figure}[htb]
\begin{center}
\includegraphics[width=0.9\textwidth]{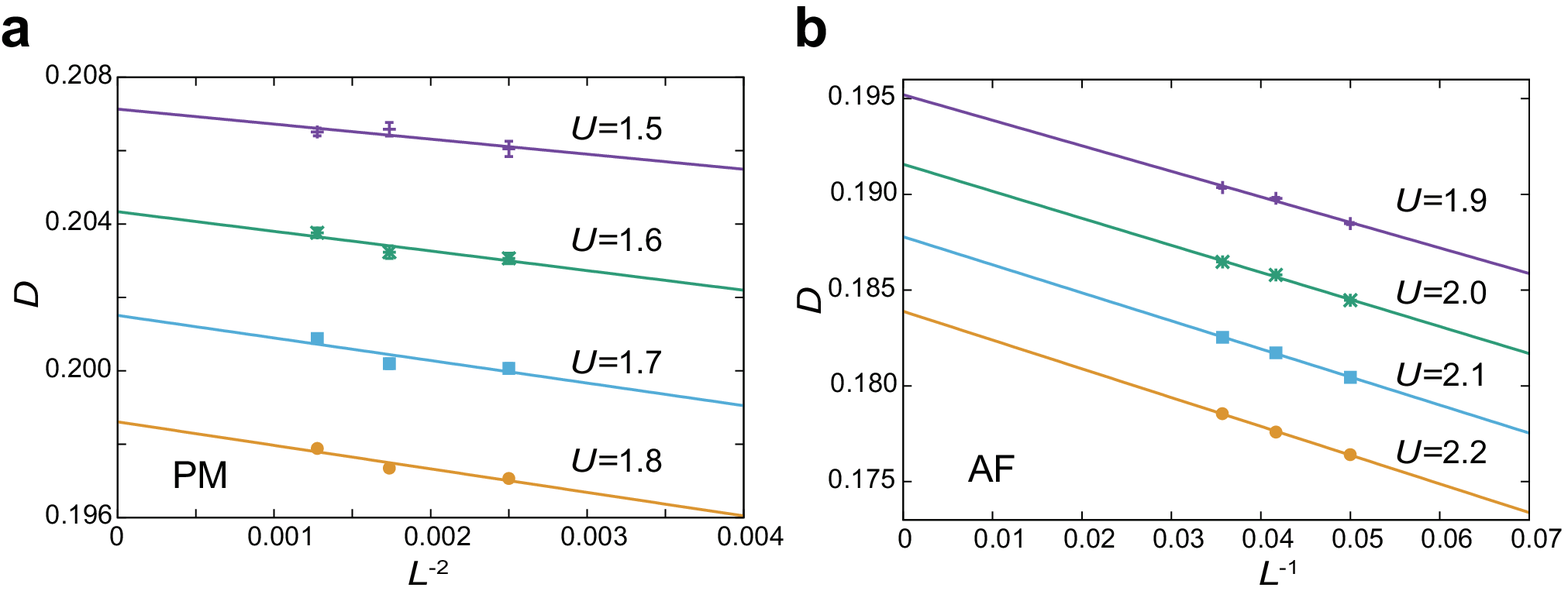}
\end{center}
\caption{\label{fig:extrap_D}
Size extrapolation of the double occupancy at  $t^{\rm MQCP}_{\perp} = 0.38$ by using Eq.~(\ref{eq:extrap_D}).
The VMC data of the double occupancy are plotted
for the metallic phase ($U=1.5, 1.6, 1.7$, and 1.8) {\bf a.} and
for the insulating phase (($U=1.9, 2.0, 2.1$, and 2.2) {\bf b.}
by using the lattices with $L=20, 24,$ and 28.
}
\end{figure}

\vskip 5mm
\noindent
{\bf H. Size extrapolation of double occupancy} \\
To analyze the ciriticality by using the double occupancy,
we perform the size extrapolations of $D$ by using the following formulae,
\begin{align}
D(L) = D_{\infty} + 
\left\{
\begin{array}{ll}
b_{\rm M}/L^2 & {\rm (metal)}\\
b_{\rm I}/L & {\rm (insulator)}\\
\end{array}
\right\} \label{eq:extrap_D}
\end{align}
where $D_{\infty}$ is the double occupancy at the thermodynamic limit and $b_{\rm M}$ ($b_{\rm I}$)
is the fitting parameter in the metallic (insulating) phase.
Examples of the fitting are shown in Fig.~\ref{fig:extrap_D}.
The error bars in Fig.~2 of the main article are determined by the square root
of the mean square error of the fitting.
%%%%%%%%%%%%%%%%%%%%%%%%%%%%%%%%%%%%%%%%%%%%%%%%%%%%
%    Figure S9
%%%%%%%%%%%%%%%%%%%%%%%%%%%%%%%%%%%%%%%%%%%%%%%%%%%%
%%%%%%%%%%%%%%%%%%%%%%%%%%%%%%%%%%%%%%%%%%%%%%%%%%%%
%%%%%%%%%%%%%%%%%%%%%%%%%%%%%%%%%%%%%%%%%%%%%%%%%%%%
\vskip 5mm
\noindent
{\bf I. Decoupling of metal-insulator and antiferromagnetic transitions} \\
 The  metal-insulator  transition  (MIT)  is  often  intertwined  with  magnetic   fluctuations.  A  phenomenology  that  will  capture  both   the  MIT  and    the   spin  degrees  of  freedom  necessitates   a  scalar  order   parameter   $\Phi(\ve{x},\tau) $   that  captures the  doublon   occupancy,  as   well  as  a  normalised 
  vector  order  parameter    $\ve{n}(\ve{x},\tau)$   that  captures  the  antiferromagnetic  fluctuations.     The  field   theory  has  to posses  a  $Z_2  \times  O(3) $  global  symmetry,  $  \Phi(\ve{x},\tau)  \rightarrow  -\Phi(\ve{x},\tau)  $    and  $\ve{n}(\ve{x},\tau) \rightarrow  O \ve{n}(\ve{x},\tau)$   with  $O$  an  orthogonal  matrix,  and   effective   Lagrangian  reads: 
\begin{equation}
       L   =   L_{\Phi}   +    L_{n}   +  L_{int}.
\end{equation}
 It  accounts  for  the  dynamics  of the  scalar field  and the   vector   field  as  well as   the  interaction  between  both of  them.  
 We  will   refrain  from  writing down  explicit  forms for  $L_{\Phi} $  and  $L_{n}$,  since  the  only  information we  need  to  assess if    $L_{int}$  is  relevant  or  not  at  criticality    are  the  scaling  dimensions of $\ve{n}(\ve{x},\tau)$ and   $\Phi (\ve{x},\tau)$.
 Assuming a single singular spatio-temporal length scale $\lambda$, and for a given dynmaical exponent $z$, we expect that the correlation function of the order parameter $\Phi$ at criticality follows
\begin{equation}
    \langle  \Phi ( \lambda \ve{x}, \lambda^z\tau)   \Phi ( \lambda \ve{x}', \lambda^z\tau')   \rangle   \propto  \frac{1}{\lambda^{2  \Delta_{\Phi}}} 
     \langle  \Phi ( \ve{x}, \tau)   \Phi ( \ve{x}', \tau')   \rangle,
\label{phiphi}
 \end{equation}
where $\Delta_{\Phi}$ is represented by the exponents of the MQCP as $2\Delta_{\Phi}=d+z+\eta-2$.
For those who are not familar with this critical scaling exponent $\Delta_{\Phi}$, see below an example of the simple $\phi^4$ model.
At  the  MQCP  our estimates are  $z\sim2$ and $\eta\sim2$  and  the  equal  time  doublon-doublon  correlation functions   are  consistent  with 
$\Delta_{\Phi}  \simeq  2$.   
A  similar  form  holds  for  the  O(3)   order  parameter  $\ve{n}$.      We  are  now  in a  position to perturbatively   understand  if  the  coupling   between the spin and doublon  degrees  of    freedom  is   irrelevant,  marginal  or  relevant.     The  most  relevant  symmetry  allowed  interaction  between  the   O(3)  and  $Z_2$  fields  reads: 
\begin{equation}
	  L_{int}   =  g \int  d^{2}\ve{x}  d \tau   \Phi(\ve{x},\tau)^2   \left(   \ve{\nabla}_{\ve{x}}   \ve{n}(\ve{x},\tau) \right)^2   +  \cdots
\end{equation}
We  note  that  due  to  the  normalization of the O(3)  order  parameter    $\Phi^2 \ve{n}^2 $    does  not  provide  a   spin-charge  coupling.     The  ellipsis    denotes  higher  order  terms  under  a scale  transformation.      Under  a  scale  transformation,  the   interaction terms  transforms  as  
\begin{equation}
g  \rightarrow   g \lambda^{ z  -  2  \Delta_{\Phi}  - 2 \Delta_{\ve{n}}} 
\end{equation}
As  mentioned  above,  we  know  that for   the MQCP  
$\Delta_{\Phi}  \simeq  2$   and  that $z\simeq 2$.   As  a  result,  and  for  any  $\Delta_{\ve{n}}  >  0$,    $g$ scales  to zero    under  successive   coarse  graining 
 scale  transformations.     The  above  provides  a  compelling  argument   supporting the notion that the  charge  and  spin  transitions   are,  
 in the  RG  sense,  independent of  each  other  at  the MQCP.

Here, we supplement the relation of the scaling exponent of the correlation defined in Eq.~(\ref{phiphi}) to the genral framework of the scaling theory in a simple examle of conventional $\phi^4$ theory for the readers who are not familiar with the scaling theory of quantum systems. The non-dimensional $\phi^4$ Hamiltonian  $H_{\phi}$ is given by
\begin{equation}
       H[\phi] = \int d^d r  [\frac{1}{2}(\nabla \phi)^2 +\frac{A}{2}\phi^2+\frac{B}{4}\phi^4 ]
\end{equation}
with coefficients $A$ and $B$. From the assumption of  a single length scale $\lambda$, this classical $\phi^4$ Hamiltonian requires the scaling of $\phi$ from the first term as 
\begin{equation}
[\phi ]=\lambda^{1-\frac{d+\eta}{2}},
\label{phi}
\end{equation}
where $\eta$ is the anomalous dimension to account for the relation of $\lambda$ and the diverging correlation length $\xi$.
From the second and third terms, we obtain similarly
$[A ]=\lambda^{-2}$ and $[B ]=\lambda^{d+\eta-4}$, respectively. When quantum dynamics is introduced,  a mapping of a $d$-dimensional quantum system to $d+z$ dimensional classical representation tells us that we need to replace $d$ with $d+z$. Here, $z$ is the dynamical exponent to represent the scaling of time scale $[\tau]=[\lambda]^z$.
From this we obtain the criticality of correlation function from the scaling of $\phi$ (Eq.(\ref{phi})) as
\begin{equation}
    \langle  \phi ( \lambda \ve{x}, \lambda^z\tau)   \phi ( \lambda \ve{x}', \lambda^z\tau')   \rangle   \propto  \frac{1}{\lambda^{2  \Delta_{\phi}}} 
     \langle  \phi ( \ve{x}, \tau)   \phi ( \ve{x}', \tau')   \rangle
\label{phiphi2}
 \end{equation}
with $2\Delta_{\phi}=d+z+\eta-2$.

%\bibliographystyle{apsrev}% Produces the bibliography via BibTeX.
%\bibliographystyle{naturemag}
%\bibliography{masterbib}% Produces the bibliography via BibTeX.

\end{document}